\documentclass[english,aps,reprint,showpacs,prb,amsmath,amssymb,floatfix,superscriptaddress,eqsecnum,twocolumn,prb]{revtex4-1}
\RequirePackage[T1]{fontenc}
\RequirePackage{times} 
\usepackage{babel}

\usepackage{amsfonts}
\usepackage{graphicx}
\usepackage{epstopdf}
\usepackage{appendix}
\usepackage{color}
\usepackage{textcomp}
\usepackage{cancel}
\usepackage[normalem]{ulem}
\usepackage{dcolumn}        
\usepackage{bm}             
\usepackage{amsmath}
\usepackage{amssymb}
\usepackage{latexsym}
\usepackage{amsbsy}
\usepackage{array}
\usepackage{setspace}
\usepackage{xcolor}
\usepackage{booktabs}
\usepackage{multirow}

\usepackage[unicode=true,
 bookmarks=true,bookmarksnumbered=false,bookmarksopen=false,
 breaklinks=false,pdfborder={0 0 1},backref=false,colorlinks=false]
 {hyperref}
\hypersetup{
 pdfauthor={Lianwei Wang}}

\usepackage{dsfont}

\begin{document}

\title{First-principles study of magnon-phonon interactions in gadolinium iron garnet}

\author{Lian-Wei Wang}
\affiliation{The Center for Advanced Quantum Studies and Department of Physics, Beijing Normal University, Beijing, 100875, China}
\affiliation{Shenzhen Institute for Quantum Science and Engineering, and Department of Physics, Southern University of Science and Technology, Shenzhen 518055, China}
\affiliation{Center for Quantum Computing, Peng Cheng Laboratory, Shenzhen 518005, China}

\author{Li-Shan Xie}
\affiliation{The Center for Advanced Quantum Studies and Department of Physics, Beijing Normal University, Beijing, 100875, China}

\author{Peng-Xiang Xu}
\affiliation{Center for Quantum Computing, Peng Cheng Laboratory, Shenzhen 518005, China}

\author{Ke Xia}
\email{kexia@bnu.edu.cn}
\affiliation{The Center for Advanced Quantum Studies and Department of Physics, Beijing Normal University, Beijing, 100875,  China}
\affiliation{Shenzhen Institute for Quantum Science and Engineering, and Department of Physics, Southern University of Science and Technology, Shenzhen 518055, China}
\affiliation{Center for Quantum Computing, Peng Cheng Laboratory, Shenzhen 518005, China}

\date{\today}

\begin{abstract}
We obtained the spin-wave spectrum based on a first-principles method of exchange constants, calculated the phonon spectrum by the first-principles phonon calculation method, and extracted the broadening of the magnon spectrum, $\Delta \omega$, induced by magnon-phonon interactions in gadolinium iron garnet (GdIG). Using the obtained exchange constants, we reproduce the experimental Curie temperature and the compensation temperature from spin models using Metropolis Monte Carlo (MC) simulations. In the lower-frequency regime, the fitted positions of the magnon-phonon dispersion crossing points are consistent with the inelastic neutron scattering experiment. We found that the $\Delta \omega$ and magnon wave vector $k$ have a similar relationship in YIG. The broadening of the acoustic spin-wave branch is proportional to $k^{2}$, while that of the YIG-like acoustic branch and the optical branch are a constant. At a specific $k$, the magnon-phonon thermalization time of $\tau_{mp}$ are approximately $10^{-9}$~s, $10^{-13}$~s, and $10^{-14}$~s for acoustic branch, YIG-like acoustic branch, and optical branch, respectively. This research provides specific and effective information for developing a clear understanding of the spin-wave mediated spin Seebeck effect and complements the lack of lattice dynamics calculations of GdIG.

\end{abstract}


\maketitle
\section{Introduction}
The collinear multi-sublattice compensated ferrimagnetic insulator gadolinium iron garnet (${\rm Gd_{3}Fe_{5}O_{12}}$, GdIG) has the same crystalline structure as YIG~\cite{GELLER195730,GELLER1959235,bertautf,yigexconstant}, only if yttrium is replaced by the magnetic rare-earth element, gadolinium.~\cite{pauthenet1959magnetic,calhoun1963anisotropy,Harris1963,Weidenbornera03267,LASSRI20113216}In comparison with YIG~\cite{dampinga,dampingb}, GdIG also has a low Gilbert damping constant of nearly $10^{-3}$,~\cite{maierflaig2017perpendicular} but has three sublattices, where the 12 Gd sublattice moments (dodecahedrals) are ferromagnetically coupled to the 8 Fe moments (octahedrals) and antiferromagnetically coupled to the 12 Fe moments (tetrahedrals),~\cite{pauthenet1959magnetic,calhoun1963anisotropy,Harris1963,yigexconstant,Weidenbornera03267,LASSRI20113216} so that GdIG has more complex spin-wave modes than YIG, which have been obtained by first-principles study of exchange interactions, indicating that the accurate calculation method can improve and compensate for the abnormality in the spin-wave spectrum caused by exchange constants.~\cite{yigexconstant,YiLiu}

GdIG has high compensation temperatures T$_{comp} =$ 286-295 K,~\cite{Soderlind2009,Tcoma,pauthenet1958spontaneous,rodic1990initial} which is close to room temperature. Recently, the heterostructures consisting of YIG~\cite{Uchida2010Spin,kenichi2010,Saitoh2013,Eiji2017} and heavy metals (FMI/NM) have been frequently used to study the spin Seebeck effect (SSE)~\cite{sseuchia,Kajiwara2010Transmission,Adachi2013} and spin Hall magnetoresistance effect (SMR).~\cite{SMRa,SMRb,SMRc}. Similar to YIG, GdIG has been frequently used to study the SSE in FMI/NM heterostructures.~\cite{sseuchia,Kajiwara2010Transmission,Adachi2013} SSE experiments have shown two sign changes of the current signal upon decreasing temperature.~\cite{Geprags2016,cramer1027} One can be explained by the inversion of the sublattice magnetizations at T$_{comp}$, where the net magnetization vanishes and the other can be attributed to the contributions of Ferrimagnetic resonance mode ($\alpha$-mode) and a gapped optical magnon mode ($\beta$-mode).~\cite{Geprags2016,cramer1027,Xiao:2010uc}
The SMR experiments shows that GdIG has a canted configuration\cite{SMRbb} and a sign change of SMR signal~\cite{SMRaa} at around T$_{comp}$. Unlike in SSE,~\cite{Geprags2016,cramer1027} the sign change of SMR is decided by the orientation of the sublattice magnetic moments associated with exchange interaction.~\cite{SMRaa}
Thus, these experiments~\cite{Geprags2016,SMRaa} indicate that multiple magnetic sublattices in a magnetically ordered system have different individual contribution and highlight the importance of the multiple spin-wave modes determined by exchange interactions. However, the microscopic mechanisms responsible for these spin current associated effect are still under investigation. A major question is whether the high-frequency magnons play an important role in the SSE, and the fitting exchange parameters used in the literature through limited experimental data~\cite{Harris1963,Plant1983,Kreisel2009} are always physically credible.

In addition to the pivotal magnon-driven~\cite{Xiao:2010uc,Uchida2010Spin,magnondrive,Adachi} effect, phonon-drag~\cite{Adachiapl,heremans} effect plays non-negligible roles in the SSE through magnon-phonon interactions,~\cite{ly18,gerrit19,Gerrit17} which play an important role in YIG based spin transport phenomena.~\cite{DaiPeng,sseuchia,Xiao:2010uc,ly18,gerrit19,Gerrit17} Thus, the understanding of the scattering process of magnon-phonon interactions is important and meaningful. In fact, the magnon-phonon thermalization (or spin-lattice relaxation) time, $\tau_{mp}$,~\cite{ly18,sanders1977effect,tmp2013} is an important parameter used to describe the magnon-phonon interactions and calculate magnon diffusion length~\cite{Xiao:2010uc,ly18}.
We have extracted the $\tau_{mp}$ ($ \sim 10^{-9}$~s) from the broadening of magnon spectrum quantitatively~\cite{YiLiu}, in good agreement with reported data~\cite{ly18,tmp2013,Agrawal2013}, however, the value is three orders of magnitude lower than the reported $\tau_{mp} \sim 10^{-6}$~s~\cite{Xiao:2010uc,sanders1977effect,tmp1960,demokritov2006bose}. For the spin-wave spectrum and phonon spectrum to aid our understanding of the magnon-phonon scattering mechanism, the temperature-dependent magnon spectrum and lattice dynamic properties of GdIG have still not been completely determined. Here, we investigate these characteristics of GdIG based on the operable and effective method used in YIG.~\cite{yigexconstant,YiLiu}

To computationally reveal the microscopic origin of SSE in these hybrid nanostructures, the magnon spectrum, phonon spectrum and magnon-phonon coupling dominant effect in GdIG will also be investigated step by step. First, we use density functional theory (DFT) technology to study the electronic structure and exchange constants, and using Metropolis Monte Carlo (MC) simulations, we obtain the Curie temperature ($T_{C}$) and compensation temperature ($T_{comp}$). Second, we obtain the spin-wave spectrum using numerical methods combined with exchange constants. Then, the phonon spectrum is studied using first-principles calculations, allowing us to extract intersecting points of magnon branch and acoustic phonon branch. In the end, we study the temperature dependence of spin moment, exchange constants, and magnon spectrum, and calculated broadening of the spin-wave spectrum of GdIG is used to extract the magnon-phonon thermalization time.

\section{Computational details and results}
\begin{figure}[htbp]
  \setlength{\abovecaptionskip}{0.5cm}
  \setlength{\belowcaptionskip}{-0.2cm}
  \centering 
 \includegraphics[trim = 1.6cm 2.6cm 2.2cm 2.2cm, clip=true, width=0.46\textwidth,scale=1.0]{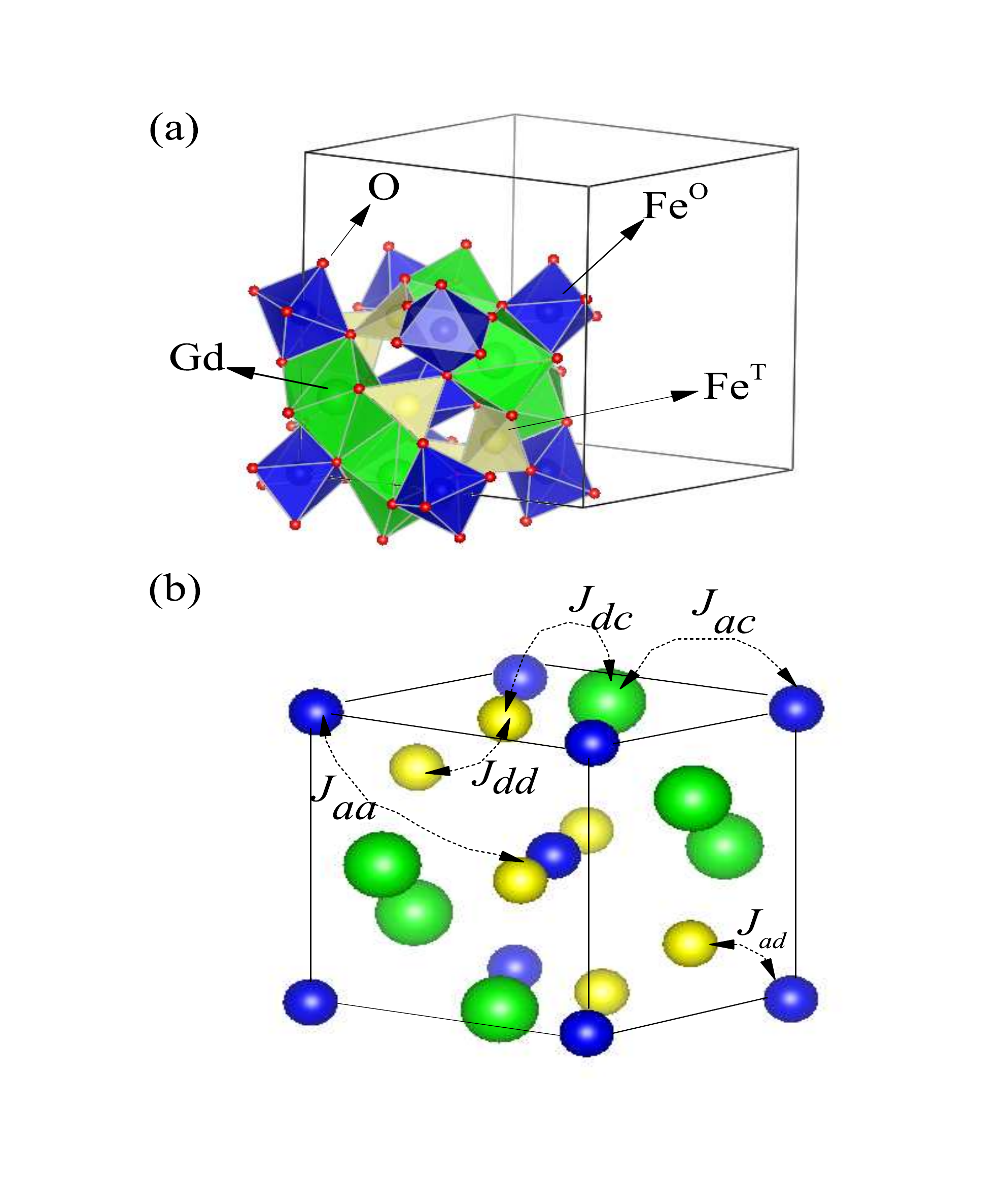} \\
 \caption{(a) 1/8 of the GdIG unit cell. The dodecahedrally coordinated Gd ions (green) occupy the 24c Wyckoff sites, the octahedrally coordinated Fe$^{{\rm O}}$ ions (blue) occupy the 16a sites, and the tetrahedrally coordinated Fe$^{{\rm T}}$ ions (yellow) occupy the 24d sites. (b) The dashed lines denote the nearest-neighbor (NN) exchange interactions. Subscripts $aa$, $dd$, $ad$, $ac$ and $dc$ stand for ${\rm Fe^{O}}$-${\rm Fe^{O}}$, ${\rm Fe^{T}}$-${\rm Fe^{T}}$, ${\rm Fe^{O}}$-${\rm Fe^{T}}$, ${\rm Fe^{O}}$-${\rm Gd}$ and $ {\rm Fe^{T}}$-Gd interactions, respectively. \label{model}}
\end{figure}

In this study, we investigate GdIG,  which belongs to the cubic centrosymmetric space group, No. 230 $Ia\overline{3}d$.~\cite{Harris1963,calhoun1963anisotropy}
The cubic cell contains eight formula units, as shown in Fig.~\ref{model}, where rare-earth gadolinium
ions occupy the 24c Wyckoff sites (green dodecahedrals), the Fe$^{{\rm O}}$ and Fe$^{{\rm T}}$ occupy the 16a sites (blue octahedrals) and 24d sites (yellow tetrahedrals), respectively, and the O ions occupy the 96h sites (red balls). The atomic sites from the experimental structural parameters (TABLE~\ref{tab:stru})~\cite{Weidenbornera03267,Harris1963,calhoun1963anisotropy} are used in the study.
\begin{table}
	\begin{ruledtabular}
	\caption{Atomic positions in the GdIG unit cell. The lattice constant is $a=12.465$~\AA.}
	\begin{center}
		\begin{tabular}{ccccc}
			& Wyckoff Position  & $x$ & $y$ & $z$ \\
		\hline
		Fe$^{\mathrm{O}}$  & 16a  & 0.0000 & 0.0000 &0.0000 \\
		Fe$^{\mathrm{T}}$  & 24d  & 0.3750 & 0.0000 & 0.2500 \\
		Gd & 24c  & 0.1250 & 0.0000 & 0.2500 \\
		O & 96h  & 0.9731 & 0.0550 & 0.1478 \\
		\end{tabular}
	\label{tab:stru}
	\end{center}
	\end{ruledtabular}
\end{table}

To calculate the electronic structure and total energy of GdIG, we use DFT, as implemented in the Vienna \textit{ab initio} simulation package (VASP).~\cite{kresse1993ab,kresse1996efficient} The electronic structure is described by the generalized gradient approximation (GGA) of the exchange correlation functional. Projector augmented wave pseudopotentials~\cite{blochl1994projector} are used. By using a 500~eV plane-wave cutoff and a $6\times6\times6$  Monkhorst-Pack $k$-point mesh we obtain results that are well converged.
	
\subsection{Electronic structure}
\begin{figure}
	\centering
	\includegraphics[width=0.48\textwidth]{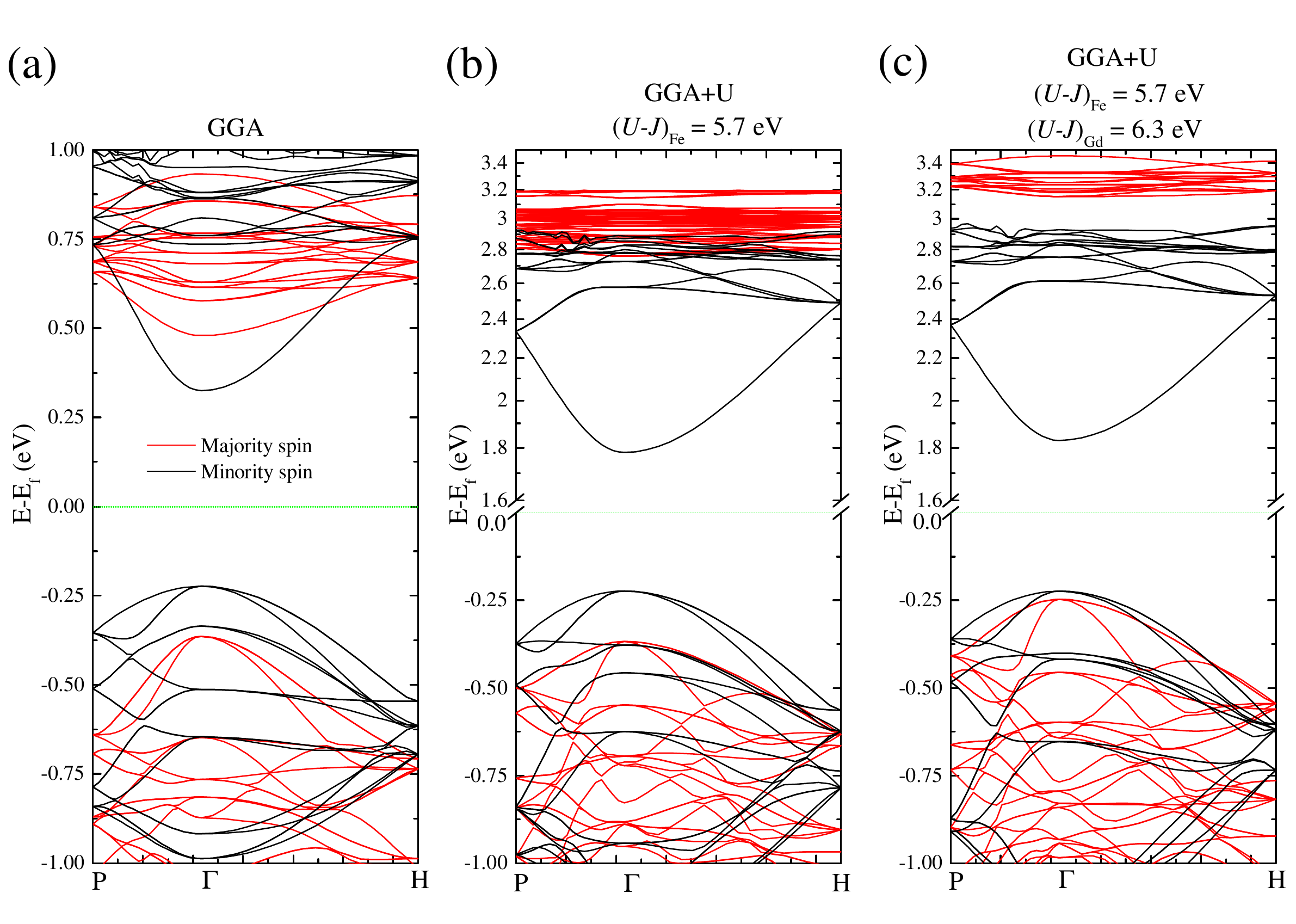}
	\caption{The energy band structure of the GdIG ground state under different calculation conditions. (a) GGA calculation results. (b) $\rm{GGA+U}$, the $d$ orbital of the $\rm{Fe}$ atom plus $U$, where the $U-J$ value is $5.7$ eV. (c) $\rm{GGA+U}$, the $d$ orbital of the $\rm{Fe}$ atom plus $U$, where the $U-J$ value is $5.7$ eV; the $f$ orbital of the Gd atom plus $U$, where the $U-J $ value is $6.3$ eV. The green lines represent 0~eV. \label{gdigbnd}}
\end{figure}
\begin{figure}
	\centering
     \includegraphics[width=0.48\textwidth,scale=2.0]{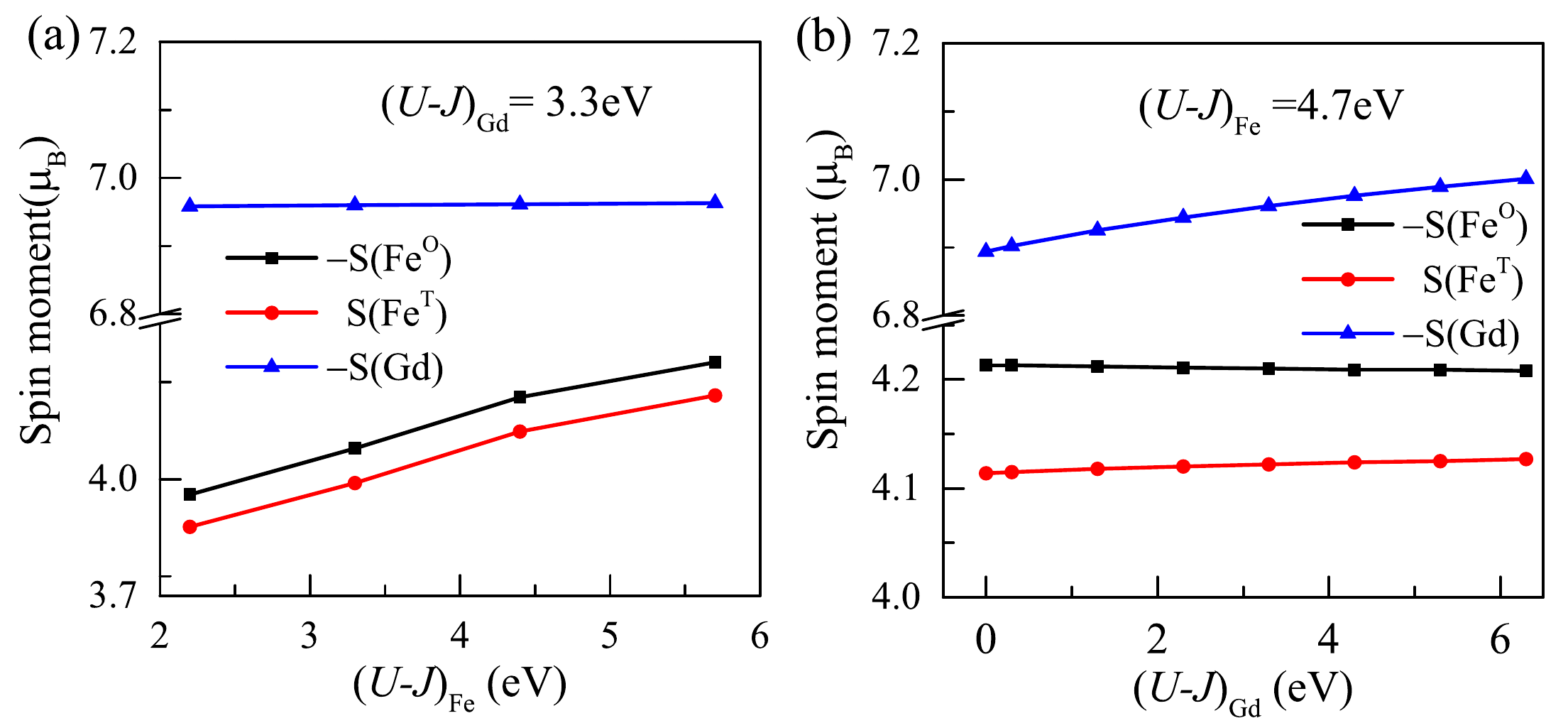}
	\caption{Variation in the spin moments of Fe and Gd ions according to different GGA+U calculations of the $f$ orbital of $\rm{Gd}$ and $d$ orbital of $\rm{Fe}$  plus $U$. (a) Variation in the fixed $(U-J)_{\rm{Gd}}=3.3$ eV calculations and (b) Variation in the fixed $(U-J)_{\rm{Gd}}=4.7$ eV calculations.\label{moment}}
\end{figure}

The calculated energy band structures of the ferrimagnetic ground-state structure, are shown in Fig.~\ref{gdigbnd}. The apparent band gap indicates the properties of the insulator. The total moment (including Fe, Gd and O ions) per formula unit is consistently 16~$\mu_B$, which is consistent with  experimental data~\cite{Heidenreich2005,LASSRI20113216}. The Fe and Gd sublattice contribute the majority of the spin moments within the unit cell.
In the DFT-GGA calculation, the spin moments of the Fe ions are $-$3.69 $\mu_{\rm B}$ for Fe$^{\rm O}$ and 3.63 $\mu_{\rm B}$ for Fe$^{\rm T}$, which are lightly larger than the computational data~\cite{LASSRI20113216}, but spin moments for the Gd and O ions are the similar values $-$6.85 $\mu_{\rm B}$ and 0.08 $\mu_{\rm B}$ respectively, and the electronic band gap is 0.55~eV, as shown in Fig.~\ref{gdigbnd}(a). And just like we did in YIG,~\cite{yigexconstant} because DFT is not good at predicting the energy gap of insulators, DFT-GGA+$U$ calculations with $U-J$ for $d$ orbital of Fe in the range of $2.2$--$5.7$~eV and $U - J$ for $f$ orbital of Gd in the range of $0.3$--$6.3$~eV are conducted to determined the Hubbard $U$ and Hund's $J$ parameters. The variation in the spin magnetic moments of different atoms under different conditions is shown in Fig.~\ref{moment}.

As shown in Figs.~\ref{gdigbnd}(b) and (c) and in Fig.~\ref{moment}, the electronic energy gap and the spin moments slightly increase with $U- J$.
For the GGA+U calculations(Fig.~\ref{gdigbnd}(b)), when the $U-J$ value for the Fe atom is constant, the band structure of the GdIG near the Fermi energy is similar to that of the YIG. When the Gd atom have $(U-J)_{\rm{Gd}}=6.3$~eV, the energy band of ${\rm Gd}$ moves up, as shown in Fig.~\ref{gdigbnd}(c). For the largest values of $U-J$, the spin moments of Fe$^{\rm O}$, Fe$^{\rm T}$, and Gd are $-$4.26~$\mu_{\rm B}$, 4.18~$\mu_{\rm B}$ and $-$7.05~$\mu_{\rm B}$, respectively, and the electric band gap is approximately 2.08~eV.  Even for the largest values of $U-J$, the moments are much smaller than expected for the pure ${\rm Fe}^{3+}$, electronic spin $S=3/2$ state $[\mu_{s}=g\sqrt{S(S+1)}=5.916 ~\mu_{B}]$ and for the pure ${\rm Gd}^{3+}$, electronic spin $S=7/2$ state $[\mu_{s}=g\sqrt{S(S+1)}=7.937 ~\mu_{B}]$. Compared with the electronic structure calculation for YIG, the results of the spin moments of Fe and the energy gap have been found to be similar.~\cite{yigexconstant}

\subsection{Exchange constants}
To obtain the five independent nearest-neighbor(NN) exchange constants, $J_{aa}$, $J_{dd}$, $J_{ad}$, $J_{ac}$ and $J_{dc}$ covering the inter- and intra-sublattice interactions, as shown in Fig.~\ref{model}(b). In TABLE~\ref{tab:etotgig}, we map ten different collinear spin configurations(SCs) a-j on the Heisenberg model without external magnetic field energy or anisotropic energy. The calculation details can be found in Ref.~\onlinecite{yigexconstant}

In the NN model, with $E_{ac}=J_{ac}S_{a}S_{c}$ and $E_{dc}=J_{dc}S_{d}S_{c}$, $E_{aa}$, $E_{dd}$, and $E_{ad}$ are just as the work in Ref.~\onlinecite{yigexconstant}, where $S_{a}$, $S_{d}$, and $S_{c}$ are the $+/-$ directions of the Fe$^{\rm O}$, Fe$^{\rm T}$ and Gd ions, the total energies, $E_{tot}$ of the Heisenberg model are determined as listed in TABLE~\ref{tab:etotgig}.
\begin{table*}[htbp]
	\caption{Comparison of the calculation of the total energies for different SCs in the NN models. The b$-$j are obtained by changing the magnetization directions of part of the magnetic ions based on the ferrimagnetic ground state SC a. $E_{tot}$ is the total energy fitting formula. $E_{cal}$ is the total energy (in units of meV) calculated via \textit{ab initio} with different $(U-J)_{\rm Gd}$ at fixed $(U-J)_{\rm Fe}= 3.4$ (in units of eV). $\Delta E$ is the difference between $E_{tol}$ and $E_{cal}$. $E_{cal}$ of  SC a is denoted as zero. }
	\begin{center}
		\begin{tabular}{ccrrrrrrrrrm{1cm}<{\centering}m{0.9cm}<{\centering}m{0.9cm}<{\centering}m{0.9cm}<{\centering}m{0.9cm}<{\centering}}
			\hline\hline
			\multirow{2}{*}{SC}    & \multirow{2}{*}{$E_{tot}$}   & \multicolumn{4}{c}{$E_{cal}$}    & \multicolumn{4}{c}{$\Delta E$} \\
			&     & 0.0 & 1.3 & 3.3 & 5.3 & 0.0 & 1.3 & 3.3 & 5.3 \\
			\hline
			a     & $E_0+32E_{aa}+24E_{dd}+48E_{ad}+48E_{ac}+24E_{dc}$ & 0.00	 & 0.00	& 0.00	& 0.00	& $-$0.01 & 0.00 & $-$0.01 	&$-$0.02 \\
			b     & $E_0+32E_{aa}+24E_{dd}-48E_{ad}-48E_{ac}+24E_{dc}$ & 3957.97 & 5036.60	& 5145.78 & 5236.59	& 0.00 	& 0.00 	& $-$0.01 &$-$0.03 \\
			c     & $E_0+32E_{aa}-24E_{dd}-48E_{ac}$  & 1729.27	& 2661.94	& 2582.27	& 2506.26	& $-$0.01 	& 0.00 	& $-$0.01 	&$-$0.03 \\
			d     & $E_0-32E_{aa}+24E_{dd}+24E_{dc}$  & 1364.90	& 2399.29	& 2454.38	& 2500.18	& $-$0.01 	& 0.00 	& $-$0.01 	&$-$0.03 \\
			e     & $E_0+32E_{aa}+24E_{dd}+48E_{ad}-48E_{ac}-24E_{dc}$ & $-$187.34	& 599.72 & 331.46 & 88.54	& $-$0.01 	& 0.00 	& $-$0.01 	&$-$0.02 \\
			f     & $E_0+32E_{aa}-24E_{dd}$  & 1770.30	& 2662.68	& 2532.24	& 2412.53	& 0.00 	    & 0.00 	& $-$0.01 	&$-$0.02 \\
			g     & $E_0+32E_{aa}+24E_{dd}+48E_{ad}$ & $-$589.27	& 299.52	& 165.63	& 43.94	    & 0.96 	& 0.34 	& 0.09 	& 0.31 \\
			h     & $E_0-32E_{aa}+24E_{dd}$  & 1807.52	& 2700.43	& 2570.38	& 2450.69	& $-$0.63 	& $-$0.54 	& $-$0.31 	& 0.00 \\
			i     & $E_0+32E_{aa}-24E_{dd}+48E_{ac}$    & 1813.35	& 2664.33	& 2482.58	& 2319.38	& $-$2.02 	& $-$0.90 	& $-$0.38 	&$-$0.60 \\
			j     & $E_0+32E_{aa}+24E_{dd}+48E_{ad}-32E_{ac}-16E_{dc}$ & $-$327.56	& 496.21	& 274.47	& 71.62	& 6.56 	& 3.56 	& 1.74 	& 2.14 \\
			\hline\hline
		\end{tabular}
	\end{center} \label{tab:etotgig}
\end{table*}
Here $E_{cal}$ are the calculated total energies for fixed $(U-J)_{\rm Fe} = 3.4 $eV and different $(U-J)_{\rm Gd}$ values relative to the ground state of SC~(a). When all or part of the magnetic moment directions of Gd atoms are flipped at $(U-J)_{\rm Gd}=0$~eV, SC~(e), (g), and (j) have lower total energies than SC~(a), which is in contrast to the experimental result, and the energy differences between these three SCs and SC~(a) decrease as $(U-J)_{\rm Gd}$ increases. Furthermore, in SC~(g), the static magnetic moment of the Gd sublattice is 0 $\mu_{B}$, and the total magnetic moment of GdIG unit molecular formula is 5 $\mu_{B}$, indicating that it is necessary to add U for Gd ions.
Through the differences of $\Delta E$ between the $E_{cal}$ and $E_{tot}$, we also find that $\Delta E$ increase when $(U-J)_{\rm Gd}$ is too small or too large. For $(U-J)_{\rm Gd}=3.3$~eV, the maximum $|\Delta E|$ is 0.63 \%, which is acceptable.
\begin{figure}
	\centering
	\includegraphics[width=0.48\textwidth]{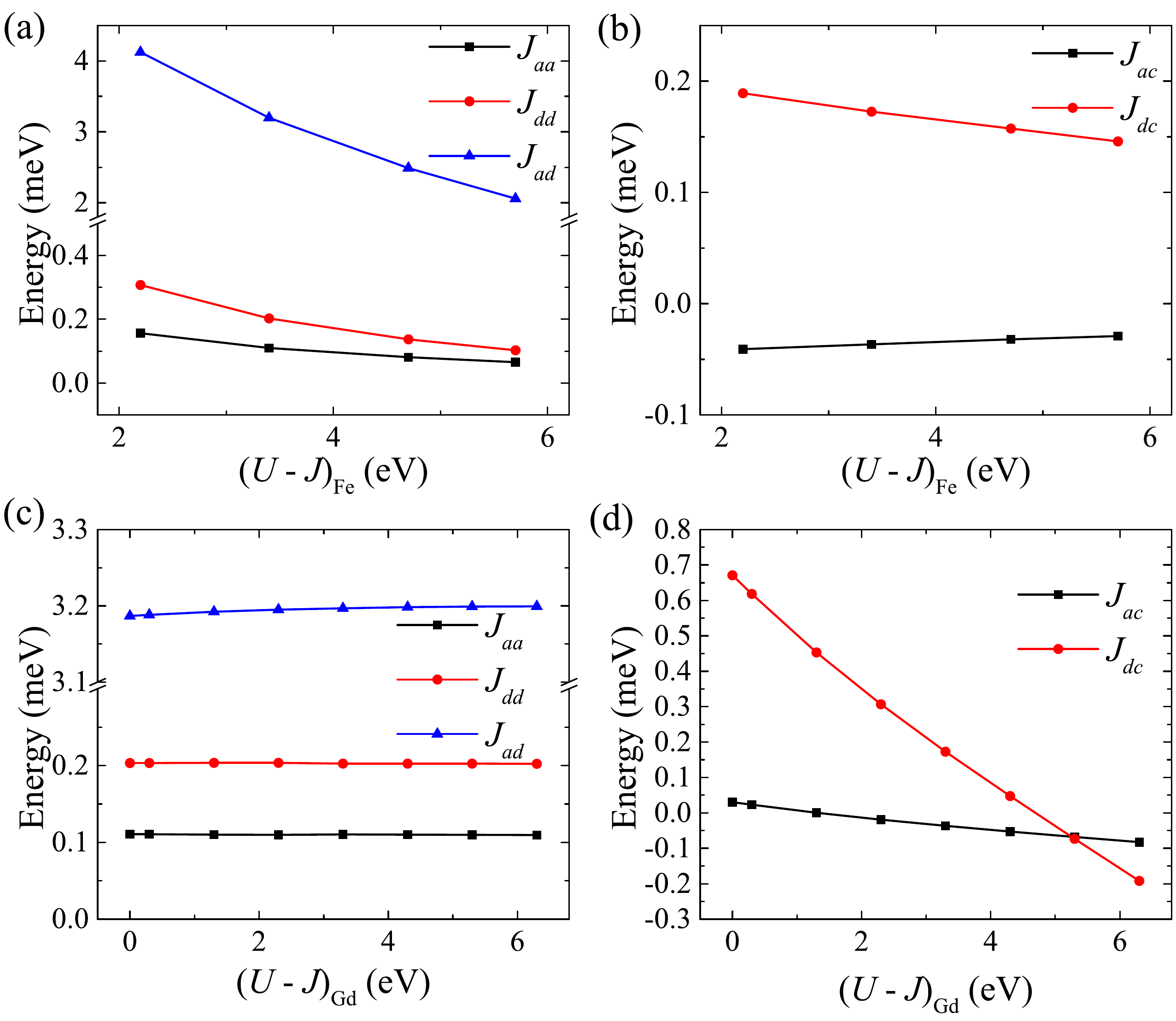}
	\caption{(a) and (b) Five different exchange interactions change with different values of $(U-J)_{\rm Fe}$ for fixed $(U-J)_{\rm Gd}=3.3$~eV. (c) and (d) Five different exchange interactions change with different values of $(U-J)_{\rm Gd}$ for fixed $(U-J)_{\rm Fe}=3.4~eV$. The negative value of the exchange constants indicates that the magnetic moment tends to be aligned in the same direction. \label{giggd4}}
\end{figure}

The exchange constants shown in Fig.~\ref{giggd4} are obtained by the least-squares of six linear equations using the SCs~a-g listed in TABLE~\ref{tab:etotgig}. SCs~h-j are selected to check whether the results are reasonable.
The exchange constants $J_{aa}$, $J_{dd}$ and $J_{ad}$ are positive (antiferromagnetic), whereas the exchange constants $J_{ac}$ and $J_{dc}$ depend on the value of $U-J$.
In Figs.~\ref{giggd4}(a) and (b), $J_{aa}$, $J_{dd}$ and $J_{ad}$ decrease as $(U-J)_{\rm Fe}$ increases when $(U-J)_{\rm Gd}$ is kept constant 3.3~eV, which is similar to the situation for YIG.~\cite{yigexconstant} The values of $J_{ad}$ are approximately 4~\% and 2~\% lower compared with the ones for YIG when $(U-J)_{\rm Fe}$ is 4.7~eV and 5.7~eV, respectively. $J_{ac}$ and $J_{dc}$ with different signs decrease slightly as $(U-J)_{\rm Fe}$ increases and for $|J_{dc}|>|J_{ac}|$.
In Fig.~\ref{giggd4}(c) and (d), when $(U-J)_{\rm Fe}$ is kept constant at 3.4~eV, $J_{aa}$ and $J_{dd}$ maintain almost the same values, whereas $J_{ad}$ increase slightly as $(U-J)_{\rm Gd}$ increases. $J_{ac}$ and $J_{dc}$ decrease to zero and then change their signs as $(U-J)_{\rm Gd}$ increases. Among all the results, $J_{ad}$ is one order of magnitude larger than the other interactions, whereas $J_{aa}$ is approximately half of $J_{dd}$ and the absolute value of $J_{ac}$ is always smaller than that of $J_{dc}$. Thus, the strong inter-sublattice exchange interaction, $J_{ad}$, dominates the other smaller energies and helps maintain the ferrimagnetic ground state of the bulk.~\cite{Harris1963,ls76,ls65} Moreover, with a change in the $(U-J)_{\rm Gd}$ value, $J_{\rm ac}$ and $J_{\rm dc}$ may change signs, which implies that it is possible to change the direction of the Gd atomic magnetic moment in the ground state.
\begin{table*}
	\caption{Exchange constants taken from the literature and our study. In calculation, $(U-J)_{{\rm Gd}} = 3.3$~eV. The unit for the interaction coefficient is meV. $(U-J)_{{\rm Fe}} = 4.7$~eV is used for comparing with the result in YIG.~\cite{yigexconstant} }
	\begin{center}
     \begin{tabular}{rrrrcclm{1cm}<{\centering}m{1cm}<{\centering}m{1cm}<{\centering}m{1cm}<{\centering}m{1cm}<{\centering}m{6.5cm}<{\centering}m{2cm}}
			\hline\hline
			\multirow{1}{*}{$J_{aa}$}  & \multirow{1}{*}{$J_{dd}$}  & \multirow{1}{*}{$J_{ad}$} & \multirow{1}{*}{$J_{ac}$} & \multirow{1}{*}{$J_{dc}$} & \multirow{1}{*}{Method} & \multirow{1}{*}{Reference} \\
			\hline
			  0.78   & 0.78  & 3.94	& 0.22	& 0.87	&Magnetization fit 	& Ref.~\cite{Harris1963}	\\
			   $-$1.05  & $-$1.47  & 3.14 	& $-$0.11	& 0.58	& Molecular field approximation 	& Ref.~\cite{ls76}	\\
			    0.56 & 1.04  &2.59 	&0.05 	& 0.16	& Molecular field approximation	& 	Ref.~\cite{ls65} \\
			    0.081 & 0.137  &2.487 	& $-$0.032	& 0.157	&\textit{ Ab initio} GGA+U ($(U-J)_{{\rm Fe}}$ = 4.7eV)	&  This paper	 \\
			     0.103& 0.185  &3.018 	& $-$0.035	& 0.170	&\textit{ Ab initio} GGA+U ($(U-J)_{{\rm Fe}}$ = 3.7eV) &  This paper	\\
			\hline\hline
		\end{tabular}
	\end{center} \label{tab:exchange}
\end{table*}

A comparison of our exchange constants with those found in prior studies is provided in TABLE~\ref{tab:exchange}. We find that different methods provide different exchange constants. Using limited experimental data, neither the magnetization fitting~\cite{Harris1963} nor the molecular field approximation~\cite{ls76,ls65} can effectively determine whether the interaction between the inter- and the intra-sublattice is ferromagnetic or antiferromagnetic coupling. Although our calculated value is smaller than the value provided in the TABLE~\ref{tab:exchange} and the obtained exchange constants between Fe atoms in GdIG are smaller than those in YIG~\cite{yigexconstant}, the relative size relationship is $J_{ad} > J_{dd} > J_{aa}$, $J_{ad} > J_{dc} > J_{ac}$. Here, we can well determine the type of exchange constants between sublattices, and use the exchange constants to obtain a reasonable experimental Curie temperature and compensation temperature. Therefore, the first-principles method of exchange constants\cite{yigexconstant,YiLiu} undoubtedly provides an effective way to calculate the interaction parameters in GdIG.

\subsection{Magnetization, Curie temperature, and compensation temperature }
\begin{figure}
	\centering
	\includegraphics[width=0.465\textwidth]{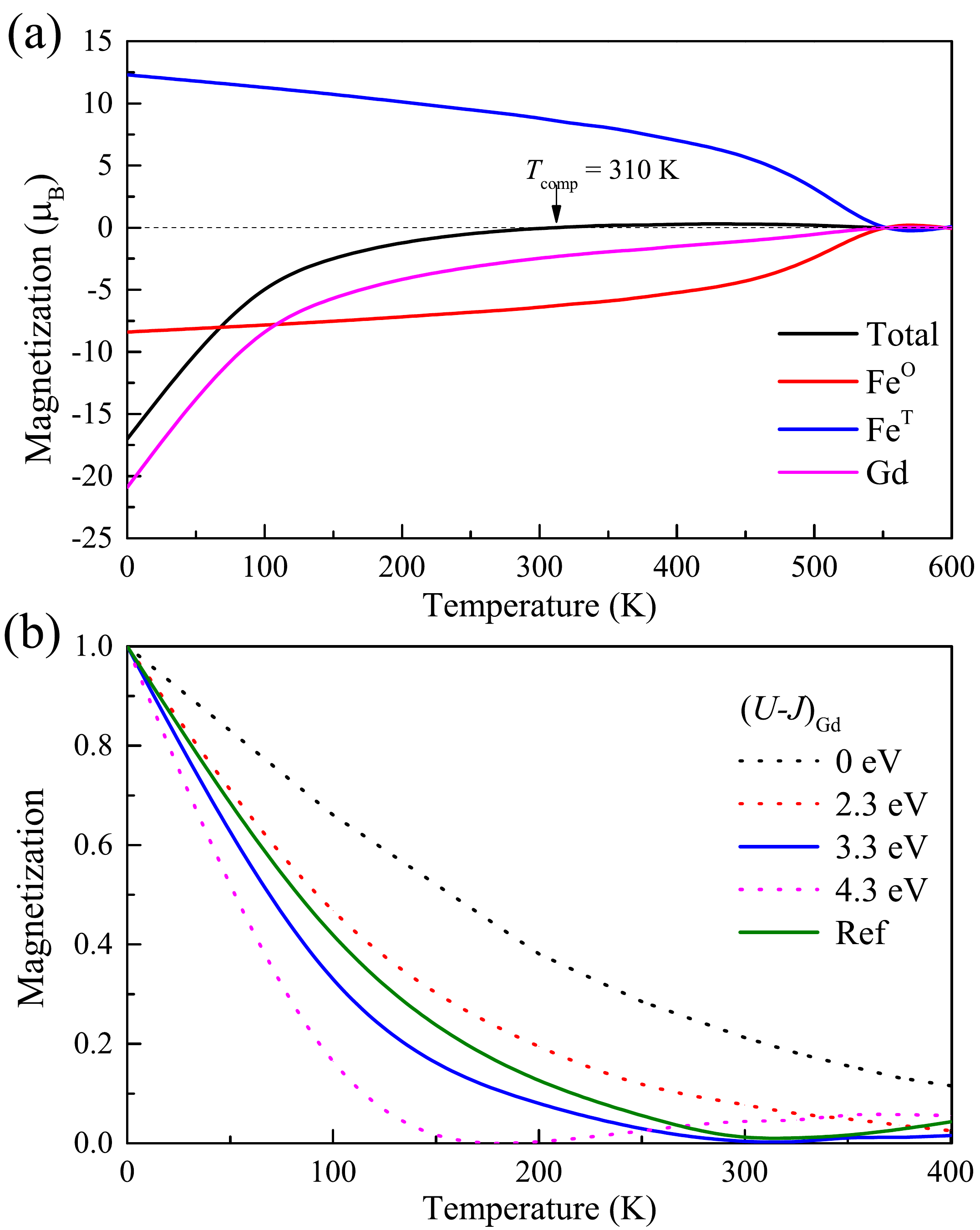}
	\caption{(a) Temperature dependence of magnetization of Fe$^{\rm O}$, Fe$^{\rm T}$, and Gd and the total magnetization of a formula unit with exchange constants fitted to the \textit{ab initio} energies for $(U-J)_{\rm Gd}=3.3$~eV and $(U-J)_{\rm Fe}=4.7$~eV. The black arrow represents the position of the compensation temperature, $T_{comp}=310$K (b) The absolute value of the total magnetization is $|M|/|M(T=0~\mathrm{K})|$ for different $(U-J)_{\rm Gd}$ at different temperatures. The reference curves (green lines) are calculated using the exchange constants in Ref.~\onlinecite{Harris1963}.
	\label{fe5u4tc}}
\end{figure}

To obtain the temperature dependence of the magnetization, Curie temperature ($T_C$), and compensation temperature ($T_{comp}$), we use the spin models by Metropolis MC simulations on a 32 $\times$ 32 $\times$ 32 supercell with a unit cell containing 32 spins under periodic boundary conditions. The computational details can be found in Ref.~\onlinecite{yigexconstant}. The results are shown in Fig.~\ref{fe5u4tc}.

With the parameters of $(U-J)_{\rm Gd}=3.3$~eV and $(U-J)_{\rm Fe}=4.7$~eV, the temperature dependence of magnetization, $M_{a}$, $M_{d}$, and $M_{c}$ of Fe$^{\rm O}$, Fe$^{\rm T}$, and Gd, respectively, and the total magnetization ($M = M_{a} + M_{b} + M_{c}$) of a formula unit are determined, as shown in Fig.~\ref{fe5u4tc}(a). The crossing point of the total magnetization curve (black) and the horizontal dash line shows that $T_{comp}$ (310~K) and $T_C$ (550~K), which are in good agreement with the experimental values of 290~K~\cite{pauthenet1958spontaneous,rodic1990initial} and 560~K~\cite{pauthenet1959magnetic,parida2008heat}, respectively.
Through the Fig.~\ref{fe5u4tc}(a), we can find that the change of the total spin moments of Fe$^{{\rm O}}$-sublattice and Fe$^{{\rm T}}$-sublattice is considerably flat. However, the total spin moment of Gd-sublattice rapidly declines with increasing temperature until approximately $200$ K; As the temperature continued to increase, owing to competition between the Gd and Fe magnetic moments, the total spin moments undergoes a transition dominated by Gd to Fe. The direction of the total magnetization changes from Gd (Fe$^{{\rm O}}$) to Fe$^{{\rm T}}$ and the value first decreases and then increases; then, $T_{comp}$ emerges~\cite{pauthenet1958spontaneous,rodic1990initial}, wherein one sign change of SSE signal appears.~\cite{Geprags2016} With a further increase in temperature, the decreasing trend of Gd-sublattice spin moments slows down. Additionally, the decreasing trend of the spin moments of Fe$^{{\rm O}}$-sublattice and Fe$^{{\rm T}}$-sublattice becomes steeper, and the total magnetic moment slowly increases and then decreases to 0~$\mu_{B}$ at transition temperature $T_C$. The temperature dependence of the magnetization of GdIG is similar to that reported in the literature.~\cite{Geprags2016}

As shown in Fig.~\ref{fe5u4tc}(b), we determine the absolute values of the total magnetization, $|M|$, normalized by its value at zero temperature, $|M(T=0~\mathrm{K})|$, for different $(U-J)_{\rm Gd}$ at the fixed $(U-J)_{\rm Fe} = 3.4 $~eV. There is no $T_{comp}$ with $(U-J)_{\rm Gd} = 0$~eV, whereas the calculated $T_{comp}$ decreases as $(U-J)_{\rm Gd}$ increases. $T_{comp}$ of the reference curve(green line) calculated using the exchange constants in Ref.~\onlinecite{Harris1963} is also approximately 310~K. Compared with Fig.~\ref{giggd4}, when the $(U-J)_{\rm Gd}$  value is small, $J_{ac}$ and $J_{dc}$ in Fig.~\ref{giggd4} are positive and $J_{ac} < J_{dc}$. Thus, the magnetization direction of Gd with Fe$^{{\rm O}}$ and Fe$^{{\rm T}}$ is anti-parallel, the latter is dominant, the ground state corresponds to SC (g) in TABLE~\ref{tab:etotgig}, and half of the Gd has an inverted magnetic moment. With increasing $(U-J)_{\rm Gd}$, $J_{ac}$ and $J_{dc}$ reverse so that Gd tended to be parallel to Fe$^{{\rm O}}$ and anti-parallel to Fe$^{{\rm T}}$; thus, the ground state corresponds to SC (a) in TABLE~\ref{tab:etotgig}. Under this condition, the $T_{comp}$ of the system decreases gradually with increasing $(U-J)_{\rm Gd}$ values. Therefore, with appropriate parameters of $(U-J)_{\rm Gd}=3.3$~eV and $(U-J)_{\rm Fe}=4.7$~eV, we can reproduce the experimental compensation temperature and transition temperature.

\subsection{Magnon spectrum}
\begin{figure}
	\centering
	\includegraphics[width=0.475\textwidth]{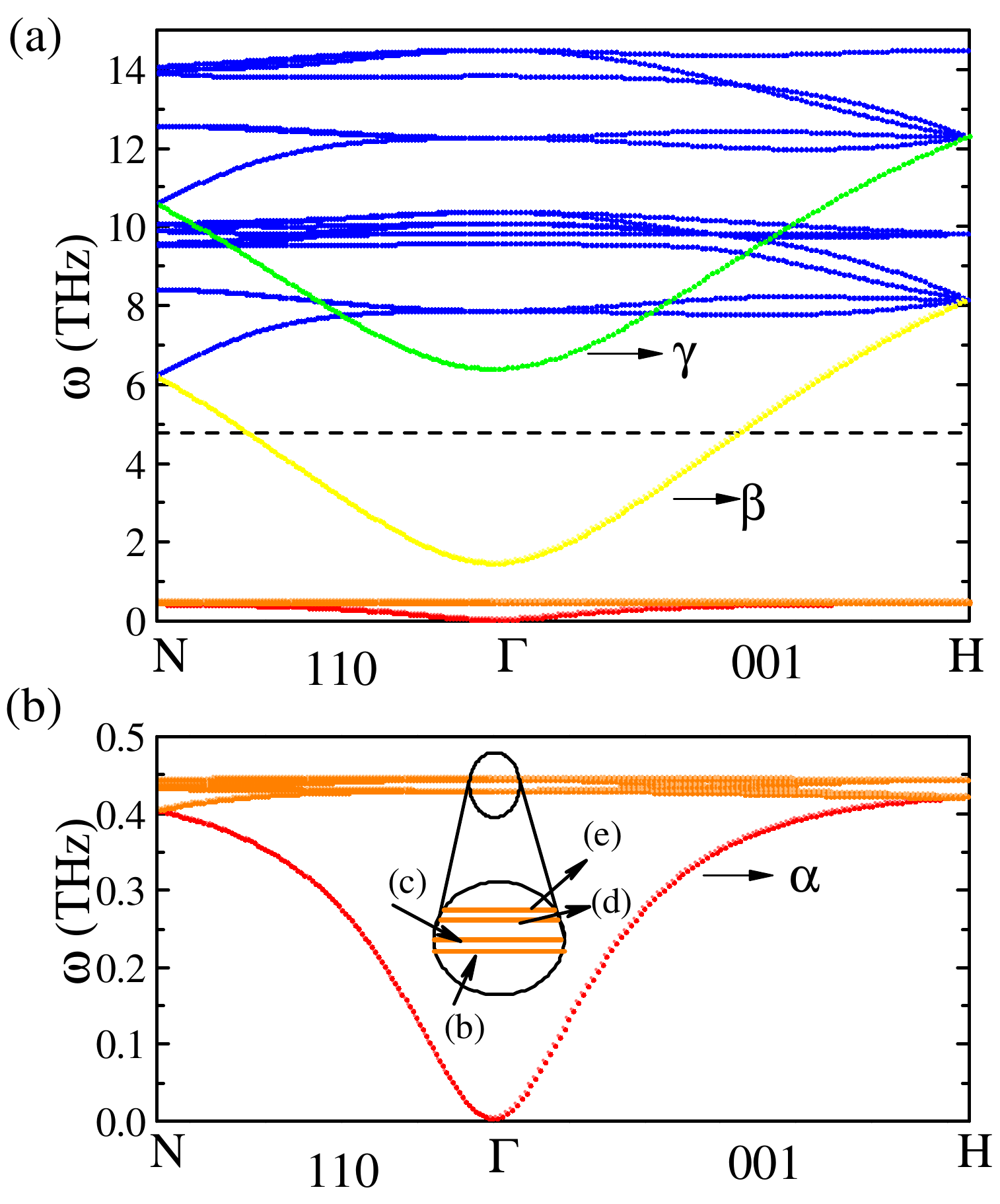}
	\caption{Spin-wave spectrum at zero temperature in the first Brillouin zone at $(U-J)_{\rm Gd}=3.3$~eV and $(U-J)_{\rm Fe}=4.7$~eV. (a) The entire spin wave spectrum. The black dash line represents the position of the lowest optical branch frequency at 4.8~THz for YIG calculated in the NN model from Ref.~\onlinecite{yigexconstant}. The notations $\alpha$ (red), $\beta$ (yellow), and $\gamma$ (green) mark the three main spin-wave modes, indicating positive, negative and positive polarization, respectively. The orange line marks the two nearly clearance modes at approximately 0.4~THz. These low-frequency optical modes are the Gd moments precession dominant. (b) The partial enlarged details of the low-frequency modes that are red- and orange- marked in (a) around 0.4~THz whereas (b), (c), (d), and (e) mark the modes at approximately 0.4 THz. The directions in the $k$-space use the standard labels for a bcc reciprocal lattice.\label{spw0}}
\end{figure}

Using the exchange interaction obtained under conditions of $(U-J)_{\rm Gd}=3.3$~eV and $(U-J)_{\rm Fe}=4.7$~eV, as shown in TABLE~\ref{tab:exchange}, we obtain the spin-wave spectrum at zero temperature, as shown in Fig.~\ref{spw0}. The details of the calculation can be found in Ref.~\onlinecite{yigexconstant}.

In Fig.~\ref{spw0}(a), the special ferrimagnetic resonance $\alpha$-mode, lower-frequency optical modes with a slight gap, YIG-like acoustic $\beta$-mode and lower-frequency optical $\gamma$-mode are marked by red, orange, yellow, and green curves. Other high optical modes are marked by blue curves.
To contrast with the lowest optical mode of YIG in Ref.~\onlinecite{yigexconstant}, a black dash line indicates the frequency. Fig.~\ref{spw0}(b) clearly shows the lower-frequency branches below 0.5 THz. The flat part (orange) has two modes at approximately $425$~GHz and two modes at approximately $440$~GHz, which are dominated by the Gd precessing.\cite{Harris1963}
The second derivative of the $\alpha$-mode at the $\Gamma$-point is $19 \times 10^{-41} ~\rm J \cdot \rm m^2$, which is approximately one quarter of the spin-wave stiffness of YIG, $D=77 \times 10^{-41} ~\rm J \cdot \rm m^2$.~\cite{yigexconstant} The $\beta$-mode around 1.4~THz has a similar parabolic branch with the acoustic branch of YIG in which the second derivative of the mode at the $\Gamma$ point is 62 $\times 10^{-41} ~\rm J \cdot \rm m^2$. The gap between $\beta$- and $\alpha$-mode at the $\Gamma$ point depends on $J_{dc}$ and $J_{ac}$ interaction,~\cite{Geprags2016} and the gap between the second parabolic lowest $\gamma$-mode and the $\beta$-mode is approximately $5$~THz, which is consistent with the conditions of the NN model in YIG.\cite{yigexconstant} As the temperature increases, the $\beta$-mode will red-shift to gain a sufficient thermal magnon population below $K_{b}T$, then the SSE signal changes sign.\cite{Geprags2016,magnondrive} However, the $\gamma$-mode will also red-shift below 6.25~THz (T$=$300 K), which may also have some indispensable effects in the SSE.

\begin{figure}
	\centering
	\includegraphics[width=0.46\textwidth]{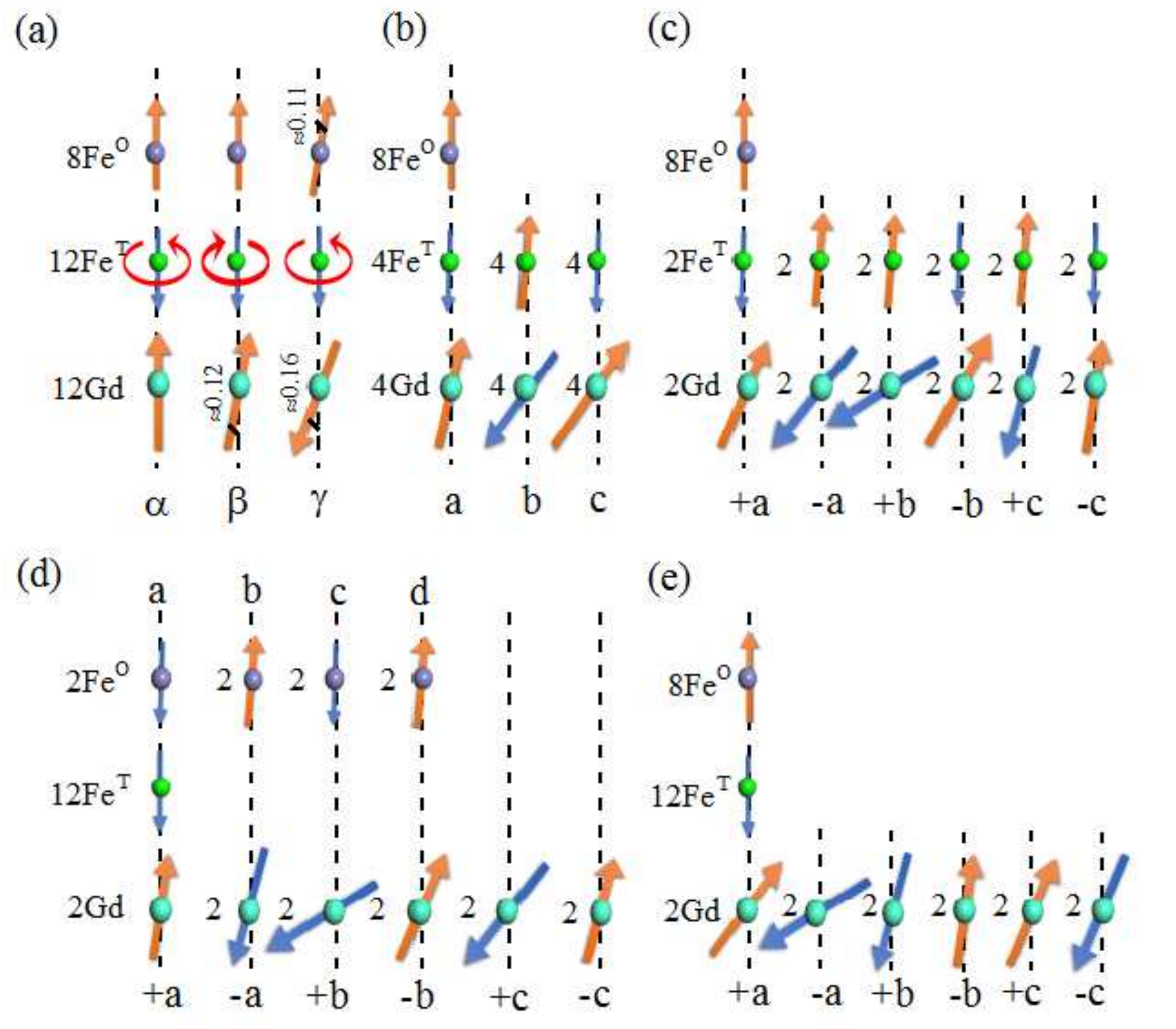}
	\caption{Precession patterns of the low-frequency modes color marked except for blue in Fig.~\ref{spw0} at the $\Gamma$-point. (a) The patterns mark the $\alpha$-, $\beta$-, and $\gamma$-mode. The red arrows represent the different chiral patterns. (b) and (c) show two patterns at approximately 425~GHz with two and three degenerated modes, respectively. (d) and (e) show two patterns at approximately 440~GHz with three and three degenerated modes, respectively. ($\pm$)a, b, c and d denote the different precession angles. The lower optical modes indicate that the Gd moments precess around the exchange field induced by Fe moments. \label{modes}}
\end{figure}

The precession patterns of these special low-frequency modes at $\Gamma$ are shown in Fig.~\ref{modes}. In Fig.~\ref{modes}(a), for the $\alpha$-mode, the magnetization of Fe$^{\rm O}$ is parallel to Gd, anti-parallel to Fe$^{\rm T}$, and near the $\Gamma$ point, where $\omega$ and $k$ satisfy the square relationship, which is similar to the acoustic mode of YIG~\cite{YiLiu}. For the $\beta$-mode, the pattern is different from the acoustic branch modes of YIG, the magnetization of 8 Fe$^{\rm O}$ ions, 12 Gd ions and 12 Fe$^{T}$ ions have different directions with respect to the $\alpha$-mode, and the magnetization of Gd has a small angle ($\approx 0.12$) with the z-axis. For the $\gamma$-mode, the magnetization of Fe$^{\rm O}$ has different directions with respect to Gd and Fe$^{\rm T}$, and the magnetization of Fe$^{\rm O}$ and Gd have small angles ($\approx 0.11$) and  ($\approx -0.16$) with respect to the z-axis, respectively. The $\alpha$-mode has different polarizations with the $\beta$-mode, but the same as $\gamma$-mode. The polarizations of $\alpha$- and $\beta$-mode switch at $T_{comp}$, which is related to the other sign change at a lower temperature in the SSE.~\cite{Geprags2016} However, the two modes induce the same sign in the detected SMR signal in a magnetic canted phase of GdIG.~\cite{SMRaa,SMRb}. Therefore, spin-wave modes need to be verified in greater detail by experiments in the future.

For the 425~GHz case, two patterns with two and three degenerated modes have Fe$^{\rm O}$ that spins lie along the z-axis, while Fe$^{\rm T}$ spins precess at small angles, as shown in Fig.~\ref{modes}(b) and (c). For the $440$~GHz case, for the two patterns with three and three degenerated modes, Fe$^{\rm T}$ spins align along the z-axis, and Fe$^{\rm O}$ spins precess at small angles or take the opposite direction as the Fe$^{\rm T}$ spins, as shown in Fig.~\ref{modes}(d) and (e). In both cases, Gd spins precess at a larger angle than the Fe spins in the exchange field of the Fe spins.~\cite{Harris1963,Geprags2016} The gap between these modes at the $\Gamma$ point and the $\beta$-mode is approximately $1$ THz, which is dominated by the interactions of Fe and Gd.
\begin{figure}
	\centering
	\includegraphics[width=0.43\textwidth,scale=2.5]{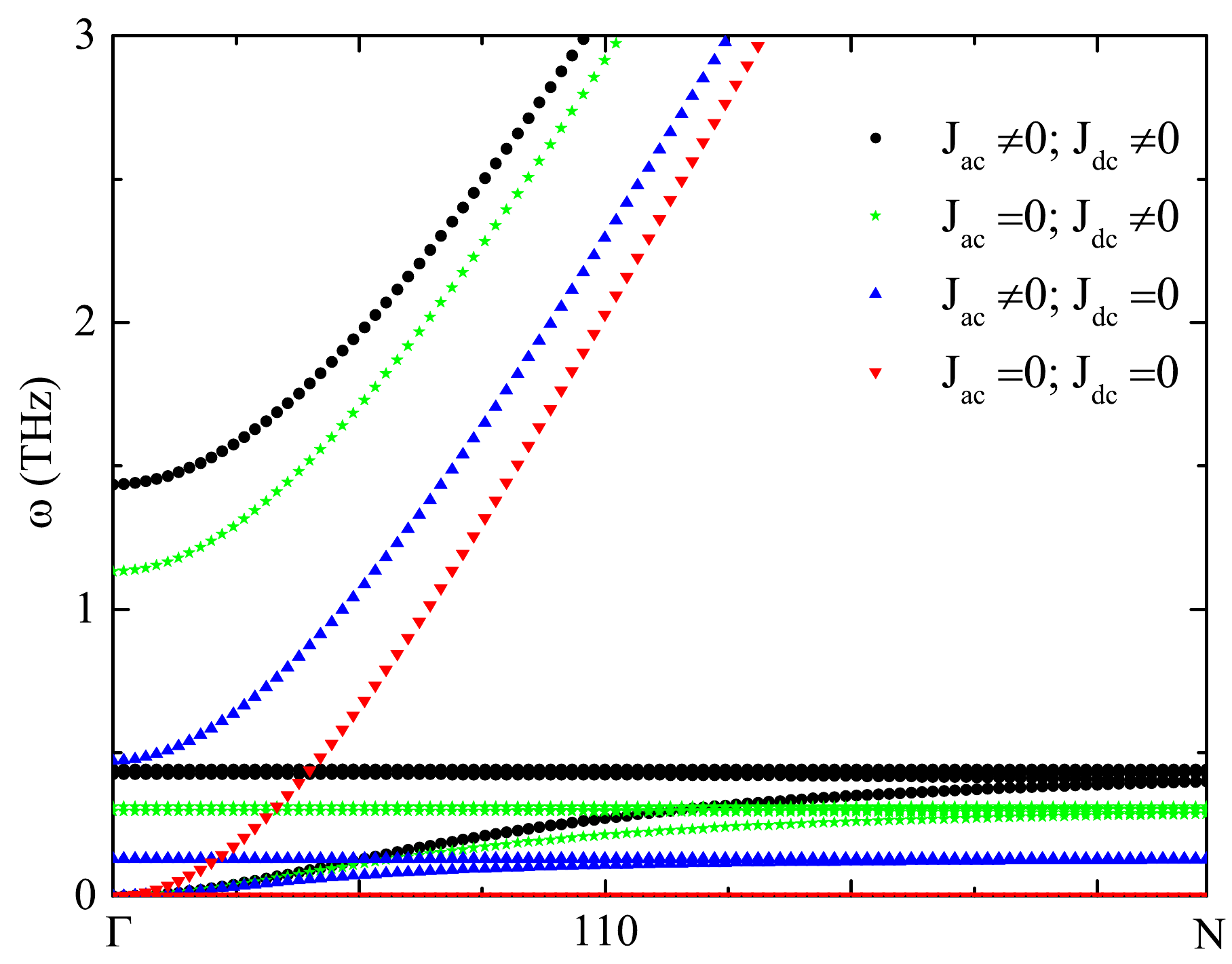}
	\caption{The spin-wave spectrum is affected by the change in exchange constants at $(U-J)_{\rm Gd}=3.3 $eV and $(U-J)_{\rm Gd}=4.7 $eV. $J_{aa}$, $J_{dd}$, and $J_{ad}$ are unchanged. For black ball curves $J_{ac}$ and $J_{dc}$ are both 0 meV; for green star curves, $J_{ac}$ is 0 meV and $J_{dc}$ is the original value; for blue triangle, $J_{dc}$ is 0 meV and $J_{ac}$ is the original value; for red triangle curves, $J_{ac}$ and $J_{dc}$ are both original values.\label{gdfe}}
\end{figure}
To show that the gap is primarily derived from the exchange interaction between the Gd atoms and Fe atoms, we fix $J_{aa}$, $J_{dd}$ and $J_{ad}$, then change $J_{ac}$ and $J_{dc}$ to show the change in spectrum along the highly symmetric direction (110), as shown in Fig.~\ref{gdfe}. We find that with the reduction of $J_{ac}$, the intersection point between the branch with the lowest frequency and the boundary of Brillouin region decreases and the band gap becomes narrower. As $J_{dc}$ decreases, the intersection point shows a more obvious reduction, the spectral lines near $0.4$ THz degenerate, and the frequency decreases. When $J_{ac}$ and $J_{dc}$ simultaneously decrease and the reduction effect is superimposed, the spectrum near $0.4$ THz completely disappears.

\subsection{Phonon spectrum }
Different \textit{ab initio} techniques and methods can be employed to calculate the phonon spectrum.\cite{dfpt,phonopy,Goel2010Lattice,sm,Maehrlein2018Dissecting} The density functional perturbation theory  method is typically used to obtain the real-space force constants of GdIG whose DFT+U ground state is used for self-consistent linear-response calculations in VASP as above. Phonon band structures, partial density of states (PDOS), total density of states (TDOS) and the phonon velocity of GdIG are investigated using the force constants via the Phonopy code.~\cite{PhysRevB.78.134106,phonopy} We also use the same calculation method to obtain the phonon spectrum of YIG, where the calculation parameters come from Ref.~\onlinecite{yigexconstant}.
\begin{figure}
	\centering
	\includegraphics[width=0.464\textwidth]{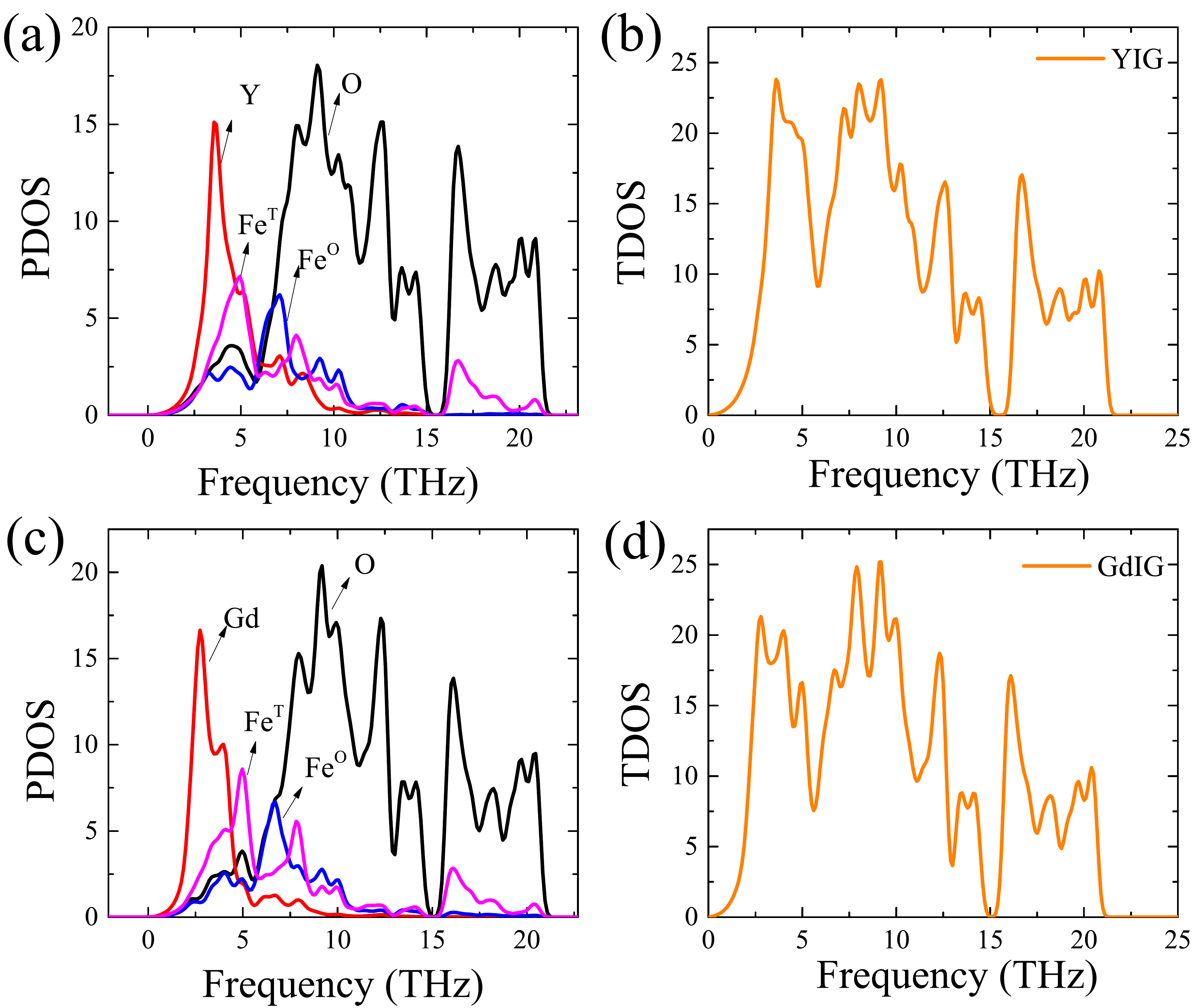}
	\caption{First-principles phonon calculation of YIG and GdIG at zero temperature. (a) Projected density of states (PDOS) for YIG. (b) Total density of states (TDOS) for YIG. (c) PDOS for GdIG. (d) TDOS for GdIG. ${\rm Fe}^{{\rm O}}$ represents the octahedral atom sites, and  ${\rm Fe}^{{\rm T}}$ represents the tetrahedral atom positions, the same representation as shown in Fig.~\ref{model}.\label{alldos}}
\end{figure}

In the calculation, we first obtain PDOS and TDOS of GdIG and YIG, as shown in Fig.~\ref{alldos}. Figs.~\ref{alldos}(a) and (c) show the PDOS for different atoms. Figs.~\ref{alldos}(b) and (d) show the TDOS. The clear results for YIG are similar to the results from Ref.~\onlinecite{sm}. The phonon gap in GdIG is approximately $2$ THz, which is consistent with the phonon spectrum for YIG. There is a difference between YIG and GdIG in the low frequency region (0-5~THz).

\begin{figure}
	\centering
	\includegraphics[width=0.465\textwidth]{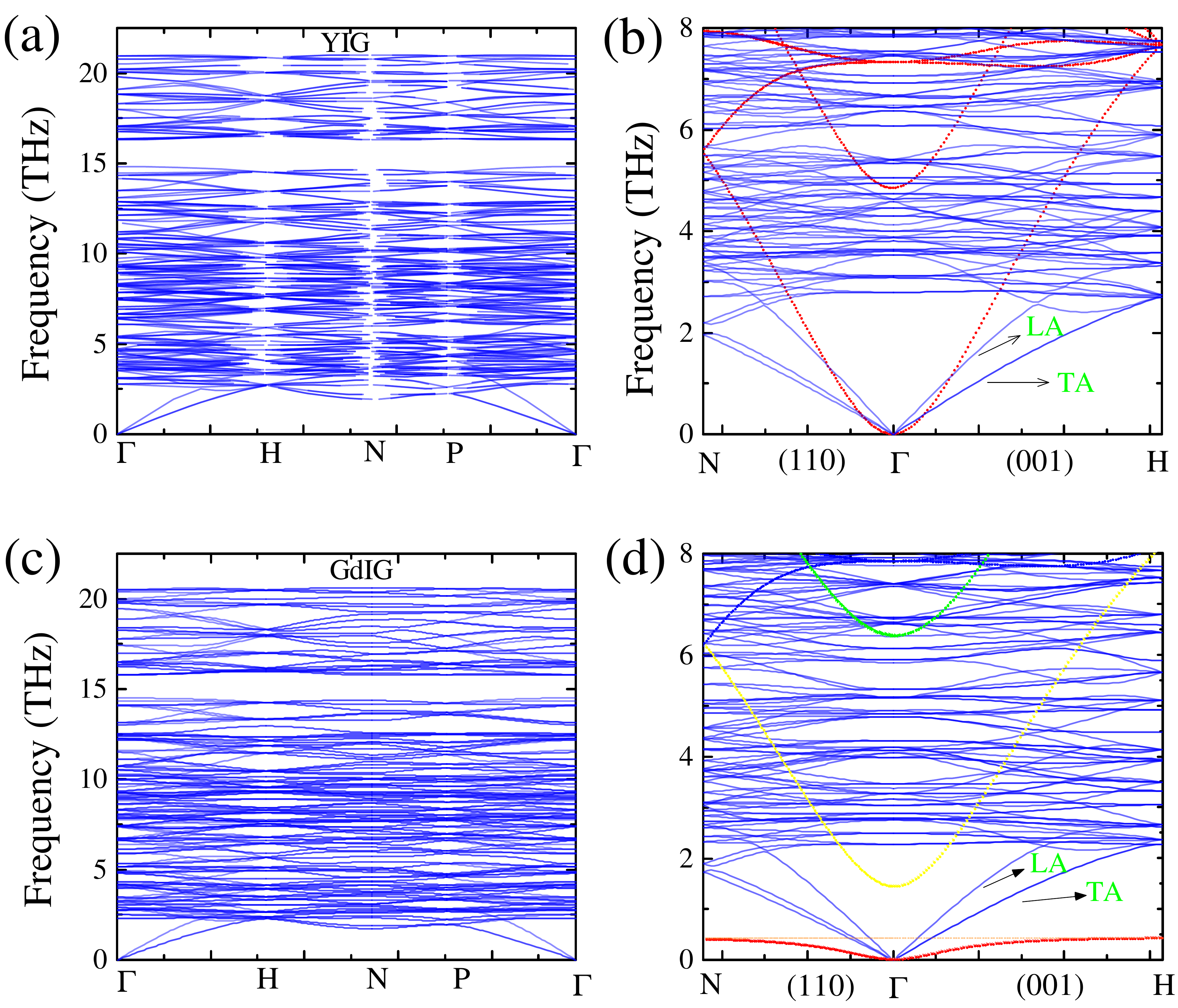}
	\caption{(a)Phonon spectrum of YIG along the $\Gamma$-H-N-P-$\Gamma$ high-symmetry lines. (b) Comparison between the phonon and spin-wave spectra along the $\Gamma$-N and $\Gamma$-H high-symmetry directions in YIG. (c) Phonon spectrum of GdIG along the same high-symmetry lines with (a). (d) Comparison between phonon and spin-wave spectra along the same directions with (b) in GdIG. The longitudinal acoustic (LA) and transverse acoustic (TA) phonons are marked on (b) and (d). The partial spin-wave spectra in (b)(red dots) are obtained from Fig.~5 in Ref.~\onlinecite{yigexconstant}, and the representations in (d) are the same as  Fig.~\ref{spw0}(a), the partial phonon spectra are extracted from (a) and (c), respectively. \label{allps}}
\end{figure}
The phonon spectrum along the path of $\Gamma$-H-N-P-$\Gamma$ in the Brillouin zone of the bcc lattice for YIG and GdIG are shown in Figs.~\ref{allps}(a) and (c), which cover $240$ phonon branches. The phonon spectrum of YIG is consistent with the results calculated by the finite-displacement method in Ref.~\onlinecite{sm}. We are interested in the low-frequency phonon branches, labeled as longitudinal acoustic (LA) and transverse acoustic (TA) phonons in Figs.~\ref{allps}(b) and (d). The frequency of the special branches shows a linear $k$ dependence in the lower frequency region and the TA modes are double degenerate. The slope of the TA(LA) phonon dispersion is presented in TABLE~\ref{vv}. For YIG, the velocity of TA (LA), $\mathbf{v}=3.8 $ kms$^{-1}$ (6.74 kms$^{-1})$, is consistent with experiment results.~\cite{DaiPeng} For GdIG, the TA (LA) velocity $\mathbf{v}=3.3$ kms$^{-1}$ (6.08 kms$^{-1})$, is almost consistent with the transverse (longitudinal) sound velocity found via experiments.~\cite{LandoltBornstein1970,spinwavegdig}

In Fig.~\ref{allps}(b) for YIG, the spin-wave acoustic branch (red) taken from the Ref.~\onlinecite{yigexconstant} has an intersecting point at a very
lower-energy approximately $1.38$ ($5.13$)~meV with the TA (LA) phonon branch. These fitting magnon-phonon intersecting points are almost consistent with experiment results.~\cite{DaiPeng}

\begin{table}
	\caption{Comparison between the  calculated and reported values of the phonon velocities for YIG and GdIG }
	\begin{center}
     \begin{tabular}{m{1.6cm}<{\centering}m{2.0cm}<{\centering}m{2.0cm}<{\centering}m{2cm}<{\centering}m{0.1cm}<{\centering}}
			\hline\hline
			\multirow{2}{*}{System}  & \multirow{1}{*}{LA velocity }  & \multirow{1}{*}{TA velocity} & \multirow{2}{*}{Source}  \\
             &    (10$^{5}$ cm/s) & (10$^{5}$ cm/s) & &                 \\
			\hline
			  YIG   & 7.200  & 3.900    & Ref.~\cite{DaiPeng}	        \\
			  YIG   & 7.209  & 3.843 	& Ref.~\cite{spinyig}	        \\
			  GdIG  & 6.500  & 3.390 	& Ref.~\cite{spinwavegdig}      \\
			  YIG   & 6.740  & 3.800 	& This paper	                \\
			  GdIG  & 6.080  & 3.300 	& This paper	                \\
			\hline\hline
		\end{tabular}
	\end{center} \label{vv}
\end{table}

In Fig.~\ref{allps}(d) for GdIG, the intersection points are more complicated than for YIG. The LA phonons and TA phonons have only one cross point with the $\alpha$-modes (red) at the $\Gamma$ point, and not intersection points with the $\beta$-modes (yellow). However, they have many more intersection points with flat lower-frequency optical branches (orange) because many multiple degenerated branches stay here as shown in Fig.~\ref{spw0}(b), such as eight crossing points along the $\Gamma$-${\rm H}$ path and eleven crossing points along the $\Gamma$-${\rm N}$ path.
We speculate that in the lower-frequency region, low-frequency lattice vibrations (phonons) can couple with magnons and there may be complicated and interesting magnon-phonon \cite{DaiPeng,gerrit19,MPxiaodi} and magnon-magnon~\cite{magnon2019} coupling effects.
The results are useful for understanding the scattering process of magnon-phonon interactions in the SSE.

\subsection{Magnon-phonon coupling}
To investigate the variation of frequency and linewidth in the spin-wave spectrum at room temperature ($T=300$~K), the temperature-induced atomic vibration is considered. The statistical mean square of the displacements, $\mathbf{u}_i$, of the $i$-th atom with its mass, $M_{i}$, are determined by the Debye model,~\cite{YiLiu}
\begin{eqnarray}\label{eq:debye}
\langle |\mathbf{u}_{i}|^{2} \rangle
 &=&\frac{9\hbar^{2}}{M_{i}k_{B}}\theta_{D}\bigg(\frac{1}{4}+\frac{T^{2}}{\theta_{D}^{2}}\int_{0}^{\frac{\theta_{D}}{T}}\frac{x}{e^{x}-1}dx\bigg),
\end{eqnarray}
where $m_{\rm Gd} = 157.25$~amu, $m_{\rm Fe} = 55.85$~amu, $m_{\rm O} = 15.99$~amu and the Debye temperature is $\theta_{\rm D} = 655.00$~K.~\cite{LASSRI20113216} Here, the change in atomic displacement does not cause significant lattice deformation. The atomic vibration displacements modeled in Eq.~\ref{eq:debye} are added to the experimental structure shown in TABLE~\ref{tab:stru}. Forty atomic configurations, which are denoted as cf01 $\sim$ cf40, respectively, are used to obtain the spin-wave spectrum. We chose the parameters of $(U-J)_{\rm Gd}=3.3$~eV and $(U-J)_{\rm Fe}=4.7$~eV for the total energy calculations because they provide reasonable $T_C$ and $T_{comp}$. The magnon-phonon relaxation time can be extracted from the broadening spin-wave spectrum. For calculation details, we refer to Ref.~\onlinecite{YiLiu}.
\begin{figure}
	\centering
	\includegraphics[trim = 2cm 2.6cm 4cm 4.6cm, clip=true, width=0.46\textwidth,scale=10.4]{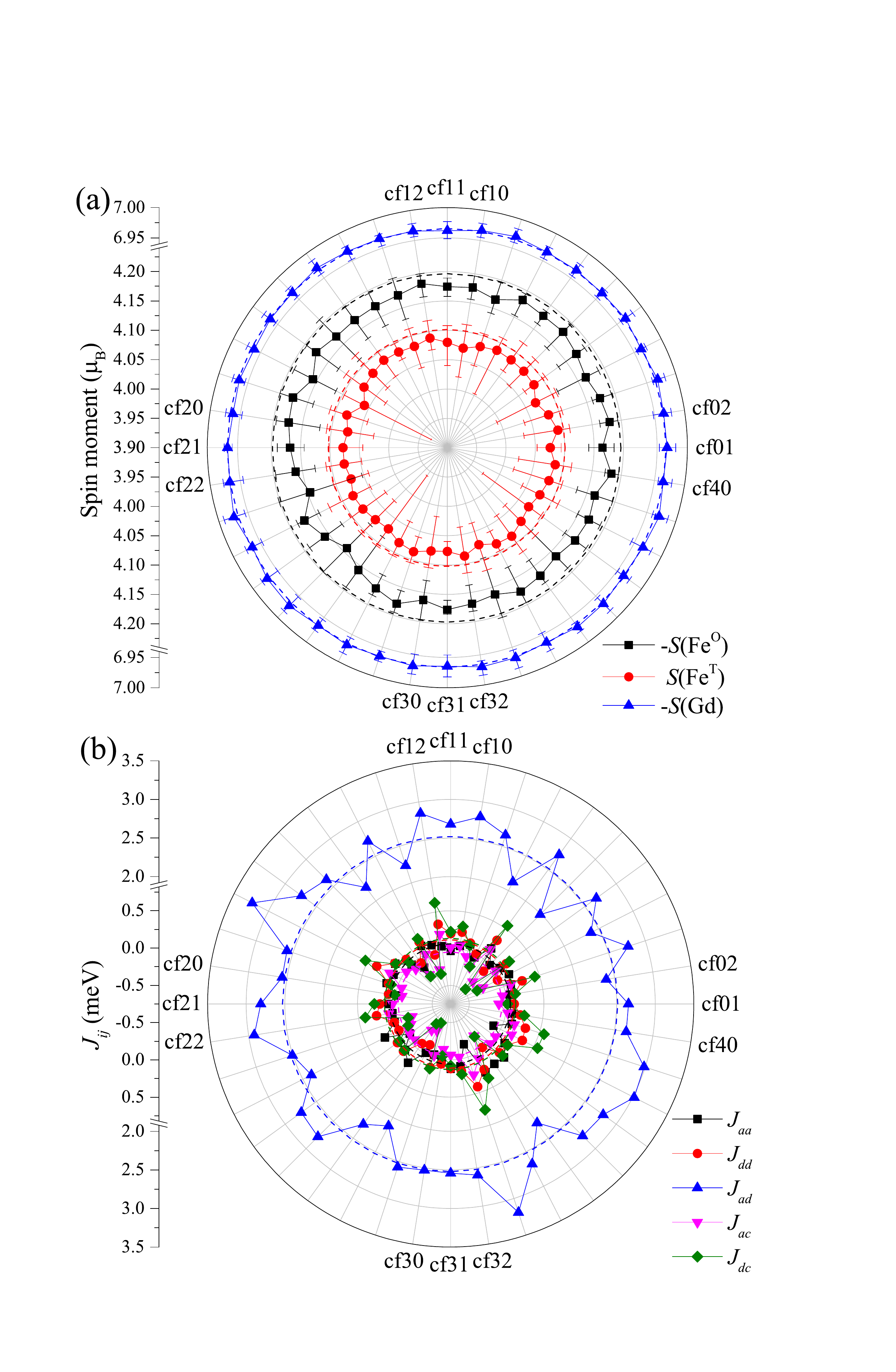}
	\caption{Spin moments and exchange constants of 40 atomic configurations (cf01-cf40). (a) Spin moments of the ions (in units of $\mu_B$) for these ground-state structures. The $-S$(Fe$^{\rm{O}}$) (black square), $S$(Fe$^{\rm{T}}$) (red dot), and $-S$(Gd) (blue triangle) curves represent different spin moments. The error bars represent the magnitude of moment change. (b) \textit{ab initio} calculation of exchange constants (in units of meV) in the NN models as like in TABLE.\ref{tab:etotgig}. The calculation parameters for $(U-J)_{\rm Gd}=3.3$~eV and $(U-J)_{\rm Fe}=4.7$~eV. The results at zero temperature are indicated by dashed lines.}\label{gdigcfmj}
\end{figure}

The spin moments of Fe and Gd ions for the ferrimagnetic ground-state structure with these 40 configurations are shown in Fig.~\ref{gdigcfmj} (a). The average moments of the $-S$(Fe$^{\rm O}$), $S$(Fe$^{\rm T}$), and $S$(Gd) ions are marked as black squares, red dots, and blue triangles, respectively. In comparison with the zero temperature values (marked as dash lines), the average moments of the Fe ions are lower for all configurations, whereas the ones for Gd ions showed no signigicant difference. The error bars denote the minimum and maximum range for each configuration. The spin moments of the Fe ions have a variation range of approximately 0.1~$\mu_{\rm B}$, which is much wider than the Gd ions at room temperature. The calculated exchange constants for each configuration are shown in Fig.~\ref{gdigcfmj}(b). The results show that the antiferromagnetic exchange constants, $J_{ad}$, still dominate. Exchange constants $J_{aa}$, $J_{dd}$, $J_{ac}$, and $J_{dc}$ may change their signs, where $J_{dc}$ has the largest variation range from $-$0.6~meV to 0.6~meV. The ground state of GdIG is still a ferrimagnetic configuration, in which the moments of the Fe$^{{\rm O}}$ atoms are arranged anti-parallel to the Fe$^{{\rm T}}$ atoms and parallel to the Gd atoms. We can see that magnon-phonon coupling can induce small fluctuation of magnetic moment and variation of exchange constants, so that the broadening of spin-spectrum can be shown.

\begin{figure}
	\centering
\includegraphics[width=0.47\textwidth,scale=5]{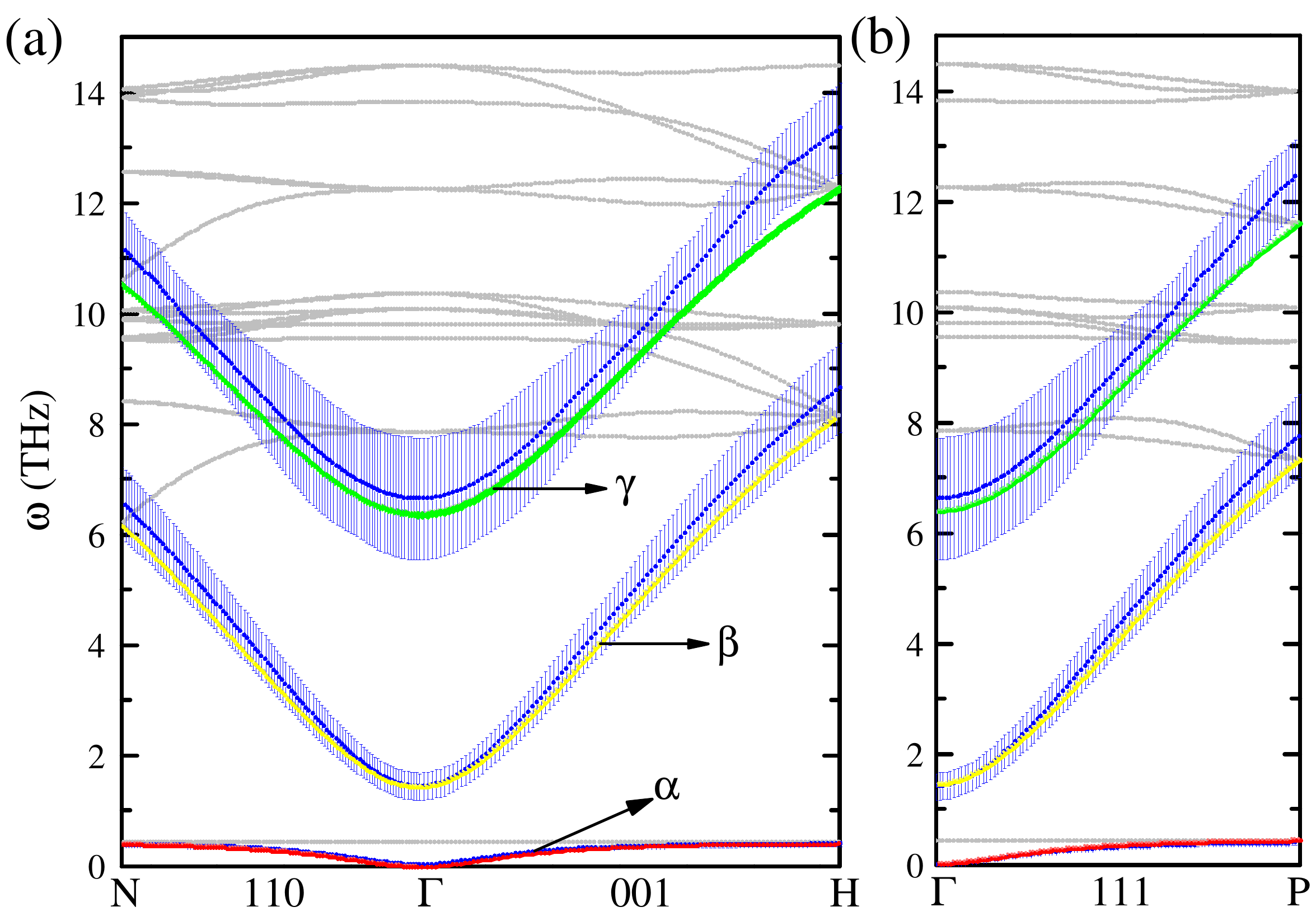}
	\caption{(a) Spin-wave spectrum in the first Brillouin zone derived from \textit{ab initio} calculations of exchange constants with $(U-J)_{\rm Gd}=3.3$~eV and $(U-J)_{\rm Fe}=4.7$~eV for 40 atomic configurations. The entire spin-wave spectrum at zero temperature is denoted using gray lines, and the picked modes for $\alpha$, $\beta$, $\gamma$ are marked in red, yellow, and green. The blue curves with error bars denote the range changes of the spectrum induced by atomic vibration. (b) Spin-wave spectrum in highly symmetric direction 111. The color means the same as in (a). The directions in the $k$-space have the standard labels for a bcc reciprocal lattice.
		\label{gdignnrt}}
\end{figure}
\begin{figure}[ht]
	\centering
	\includegraphics[width=0.48\textwidth]{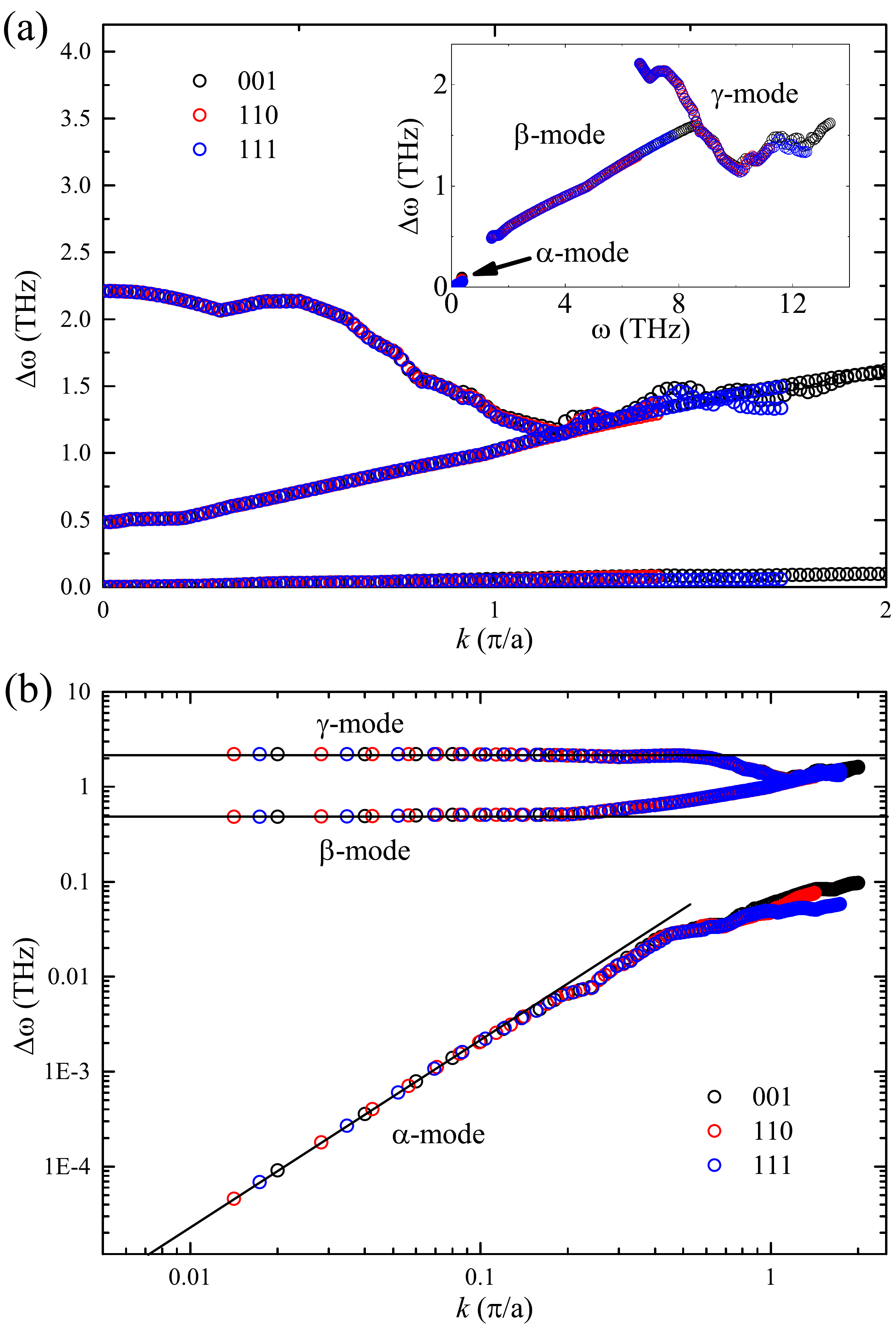}
	\caption{(a) Calculated broadening of the spin-wave spectrum of GdIG, $\Delta \omega$, at room temperature as a function of $k$. Inset: $\Delta \omega$ replotted as a function of the spin-wave frequency, $\omega$. (b) $\Delta \omega(k)$ replotted on a log-log scale. The black lines indicate a constant $\Delta \omega$  and a quadratic dependence on $k$ for the optical branch ($\beta$-, $\gamma$- modes) and the acoustic branch ($\alpha$-mode), as shown in Figs. ~\ref{spw0} and \ref{gdignnrt}, respectively.
	\label{gigdeltao}}
\end{figure}

As shown in Fig.~\ref{gdignnrt}, at room temperature, spin-wave modes are plotted as the blue curves with error bars governed by the NN exchange constants in Fig.\ref{gdigcfmj}(b). At zero temperature, the lowest frequency $\alpha$-mode and two slightly higher frequency parabolic $\beta$- and $\gamma$-mode are shown by the red, yellow, and green curves, respectively, which is the same as Fig.\ref{spw0}. Other modes are marked by gray curves.
We can see that the blue curves can superimpose with other modes and show a significant spread in energy at room temperature.
For the $\alpha$-modes, the frequency of different phonon configurations is nearly the same as red curves in different directions, and the spectral line had a slightly larger distribution range at the Brillouin zone boundary.
For the $\beta$-modes, the spectral lines distribute around the yellow curves, and the distribution range increase as the $k$ value increases in all directions. Compared with the acoustic branch of YIG,~\cite{YiLiu} the spectrum shows a smaller distribution range.
For the $\gamma$-modes, the spectral lines also distribute around green curves and disperse larger than the $\beta$-mode. However, the distribution in all directions decreases then increases with increasing $k$, which is not the case for the YIG.~\cite{YiLiu}. So the spin-wave spectrum using the phonon configurations at room temperature shows a noticeable broadening.
As shown in Fig.~\ref{gigdeltao}, the broadening of the spectrum, $\Delta\omega$, for the $\alpha$-, $\beta$-, and $\gamma$-modes are extracted from $40$ room-temperature configurations by using the method in Ref.~\onlinecite{YiLiu}.

In Fig.~\ref{gigdeltao}(a), $\Delta \omega$ has a strong dependence on the value of $k$. When the $k$-value is small, $\Delta \omega$  in the three high symmetry directions are very close to each other, and have different trends with increasing $k$.
For the $\alpha$-mode, $\Delta \omega$ increase slowly with increasing $k$, but in the (111) direction, there is a slight decrease when the $k$ value approaches the Brillouin zone boundary. Compared with the three directions, we find the relationship of $\Delta \omega(001) > \Delta \omega (110)> \Delta \omega(111)$  in the region of $k > \pi/a$.
For the $\beta$-mode, $\Delta \omega$ increases with increasing $k$. Upon comparing the three directions, we find the relationship of $\Delta \omega(001) > \Delta \omega (111)> \Delta \omega(110)$ in the region of $k > \pi/a$.
For the $\gamma$-mode, $\Delta \omega$ decrease as $k$ increase; however, in the (001) direction, there is a small increase when $k$ approaches the Brillouin zone boundary. Upon comparing the three directions, we find the relationship of $\Delta \omega(001) > \Delta \omega (110)> \Delta \omega(111)$ in the region of $k > \pi/a$.
The trend for the three modes can also be obtained from the inset in Fig.~\ref{gigdeltao}(a), where the curves in each direction are almost exactly the same, indicating that the anisotropy plays a negligible
role in the broadening $\Delta \omega$. Combined with Fig.\ref{modes}(a), we find that the $\alpha$- and $\gamma$- modes have the same positive polarization direction, and the trend of broadening is consistent as the wave vector changes in different directions. However, the $\beta$- mode has a negative polarization direction, and the trend is different.

Using the broadening $\Delta \omega$ of the spin-wave spectra at room temperature and the uncertainty relationship of $\Delta \omega \cdot \tau_{mp} = \hbar $, we calculate the magnon-phonon thermalization time, $\tau_{mp}$ or spin-lattice relaxation time to explore the magnon-phonon interactions. In Fig.~\ref{gigdeltao}(b), $\Delta\omega$ is replotted on a logarithmic scale for observing the asymptotic behavior in the long-wavelength region, where we find a quadratic dependence on $k$ of $\Delta\omega$ for the $\alpha$- modes and constants $\Delta \omega^{\beta} = 2.07$ meV and $\Delta \omega^{\gamma} = 9.14 $ meV for the $\beta$- and $\gamma$- modes, respectively, corresponding to $\tau_{mp}^{\beta}= 3.18\times10^{-13}$ s and $\tau_{mp}^{\gamma}= 7.19\times10^{-14}$~s as illustrated by the black solid lines.

As shown in Fig.~\ref{gdigcfmj}, the lattice vibrations can induce fluctuations of magnetic moment and the exchange constants. Additionally, the phonon-induced fluctuation of the exchange constants has an obvious effect on the magnon spectrum and can induce broadening of the spin-wave spectrum at room temperature (as shown in Fig.~\ref{gdignnrt}). At the long-wavelength limit ($k$ $\rightarrow$ 0), the acoustic phonon represents the centroid motion of atoms in the same unit cell, so that the change in atomic displacement caused by the temperature has little effect on the lattice; so for the $\alpha$-mode, lattice vibration induced spin-wave broadening is approximately zero. In addition, the decay rate of the spin-wave is found to be proportional to the square of $k$ at the long-wavelength limit, as shown in the hydrodynamic theory for spin-wave.~\cite{halperin,halperina} Thus $\tau_{mp}$ is proportional to $k^{-2}$ for the acoustic $\alpha$-mode. For $\beta$- and $\gamma$- modes, as the optical phonon represents the reverse motion of the positive and negative ions in the unit cell, the temperature causes the fluctuation of the average displacement of atoms, which can induce a constant spin-wave broadening, so $\tau_{mp}$ is constant for the optical modes.

To compare with YIG, we also chose a specific wave vector, $k=5.67\times 10^{5}$ ~${\rm cm^{-1}}$, from Ref.~\onlinecite{YiLiu,Agrawal2013}, and values for the $\Delta\omega$ of three modes are $6.49\times10^{-5}$ ~THz, $4.86\times10^{-1}$ THz, and $1.70$ THz. We obtain $\tau_{mp}= 2.45\times10^{-9}$~s, $3.27\times10^{-13}$~s, and $9.36\times10^{-14}$~s for the $\alpha$-, $\beta$-, and $\gamma$-modes, respectively, which are approximately $4.3$ times, $0.6\times10^{-3}$ times, and $0.1$ times the values for the acoustic branch and lowest-frequency optical branch of YIG.
As shown in Fig.~\ref{allps}(d), for the YIG-like $\beta$-mode, the sufficient density of state of the phonons can induce larger magnon-phonon scattering rate in the long-wavelength region so that the magnon-phonon thermalization time $\tau_{mp}$ is rather small, which is similar to the case of YIG.~\cite{YiLiu}. For the optical $\gamma$-mode, it also has a relatively high frequency, where the phonons have a large density of state so that the magnon-phonon scattering rate is quite large, which can result in a smaller magnon-phonon thermalization time.

\section{conclusion}
In conclusion, we investigate the NN exchange interaction coefficient using a more reliable and accurate method, which has been applied to YIG. We obtained the Curie temperature and magnetic compensation temperature that matched the experiment well. We found that the spin-wave spectrum obtained by numerical methods using the exchange constants can explain the experimental phenomena in SSE well. We reveal the spin-wave precession mode in the low frequency region, which indicates that the acoustic branch $\alpha$-modes and YIG-like optical branch $\beta$-modes have different chiral characteristics, but the same as the lower optical $\gamma$-modes. A first-principles phonon calculation method was used to obtain the phonon spectrum of GdIG and YIG at zero temperature. We reproduce the fitting intersecting point of the spin-wave and phonon branches(LA, TA) that are in good agreement with experiment results in the very low-energy region. We discuss the interaction between magnons and phonons in GdIG by introducing temperature-dependent lattice shifts. Three special spin-wave modes ($\alpha$, $\beta$, and $\gamma$) are found to exhibit different broadening of the spin-wave spectrum, $\Delta \omega$ of GdIG. In a small wave vector region, the $\Delta \omega$ of the $\alpha$- modes have a square relationship with wave vector k ($\Delta \omega \sim k^{2}$). For the $\beta$- modes, the $\Delta \omega$ are nearly a constant, which is similar to the lower optical branch of YIG.~\cite{YiLiu} A higher optical branch $\gamma$-mode also exists below $K_{b}T \sim 6.25$~THz at room temperature, which may play an indispensable role in magnon-phonon coupling, and the $\Delta \omega$ has also a constant relationship with $k$. At a particular wave vector, the magnon-phonon thermalization time, $\tau_{mp}$, for these branches at room temperature is also different from that of YIG. $\tau_{mp}\sim 10^{-9}$~s for the $\alpha$-mode is bigger than the acoustic branch of YIG, the $\tau_{mp}$ of the $\beta$- and $\gamma$-mode ($\sim 10^{-13}$ and $\sim 10^{-14}$~s) are smaller than the acoustic branch and lower optical branch of YIG, respectively. The magnon-phonon coupling effect may play more central role in higher spin-wave modes compared with lower modes.

Additionally, we also do \textit{ab initio} phonon calculations using the finite-displacement method in the packages VASP,~\cite{kresse1993ab,kresse1996efficient} ABACUS,~\cite{ABACUS} and QUANTUM ESPRESSO(QE) package~\cite{qe2019} combined with Phononpy~\cite{PhysRevB.78.134106,phonopy} to obtain the phonon spectrum of YIG and GdIG, and the results are consistent with those presented in this paper (not shown here). A well-known problem with most of the theories of magnon-phonon coupling is that they do not take into account the magnon-magnon coupling or magnon-phonon coupling directly. Thus we aim to develop a set of first-principles calculations in the future to include full interactions to study magnon transport properties and lattice dynamics.

\begin{acknowledgments}
The authors would like to thank Li-Xin He and Xiao-Hui Liu for helpful discussions about ABACUS phonon calculation, and thank Joseph Barker for providing atomistic spin dynamics simulation program to calculate Curie temperature. This work was financially supported by National Key Research and Development Program of China (Grant No. 2017YFA0303300 and  2018YFB0407601) and the National Natural Science Foundation of China (Grants No.61774017, No.11734004, and No.21421003). We also acknowledge the National Supercomputer Center in Guangzhou (Tianhe II) for providing the computing resources.
\end{acknowledgments}


\end{document}